%% file: main.tex
\documentclass{article}

\PassOptionsToPackage{numbers, compress}{natbib}
\usepackage[preprint]{neurips_2026}

\usepackage[utf8]{inputenc}
\usepackage[T1]{fontenc}
\usepackage{hyperref}
\usepackage{url}
\usepackage{booktabs}
\usepackage{nicefrac}
\usepackage{microtype}
\usepackage{graphicx}
\PassOptionsToPackage{table}{xcolor}
\usepackage[most]{tcolorbox}
\usepackage{pifont}
\usepackage{enumitem}
\usepackage{tikz}
\usepackage{algorithm} 
\usepackage{algorithm}      
\usepackage{algpseudocode}  
\usepackage{booktabs, longtable}
\usepackage{fontawesome5}
\usepackage{wrapfig}
\usetikzlibrary{arrows.meta, positioning, fit, backgrounds, calc}

\title{SAGE: Scalable Agentic Grounded Evaluation \\ for Crop Disease Diagnosis}

\author{
  \textbf{Muhammad Arbab Arshad}$^{1,*}$ \quad \textbf{Tirtho Roy}$^{1,*}$ \quad \textbf{Yanben Shen}$^{1}$ \quad \textbf{Dinakaran Elango}$^{1}$ \\
  \textbf{Shivani Chiranjeevi}$^{1}$ \quad \textbf{Asheesh K.\ Singh}$^{1}$ \quad \textbf{Baskar Ganapathysubramanian}$^{1}$ \quad \textbf{Chinmay Hegde}$^{2}$ \\
  \textbf{Arti Singh}$^{1}$ \quad \textbf{Soumik Sarkar}$^{1,\dagger}$ \\[2ex]
  \small $^1$Iowa State University, USA \quad $^2$New York University, USA \\
  \small $^*$Equal contribution \quad $^\dagger$Corresponding author: \texttt{soumiks@iastate.edu} \\[1ex]
}
\begin{document}
\maketitle

\begin{abstract}
Plant disease diagnosis is critical for food security, yet training disease-recognition models that generalize across crops, pathogens, and field conditions remains challenging because labeled disease images are far less abundant and standardized than data for other biotic stresses such as insects or weeds. Frontier vision-language models offer new opportunities through improved visual reasoning, but they still struggle with fine-grained disease identification due to the lack of structured, crop-specific symptom knowledge. To address this gap, we curate the largest plant disease image--symptom dataset to date, covering 335 crops, 1{,}251 disease classes, and approximately 839K images, designed to support training-free, agentic disease prediction. A scalable automated pipeline generates source-grounded symptom descriptions in which each claim is linked to a verbatim web quote; domain experts validate sampled crops and reconcile disease-name variants across sources. As a baseline, we introduce an autonomous visual reasoning agent that identifies anatomical context, narrows candidate diseases using symptom knowledge, sequentially compares reference images, and produces a fully explainable reasoning trace. Incorporating symptom knowledge improves accuracy by 16.2 percentage points on average at the full reference budget, with consistent gains across all four evaluation crops. Because the framework only requires crop-specific reference images and symptom knowledge, it can be extended to new crops without retraining, while the agentic baseline can directly benefit from future improvements in foundation model capabilities. Dataset and code are available at: \url{https://sage-dataset.github.io/}.
\end{abstract}

\section{Introduction}
\label{sec:intro}

Infectious diseases reduce potential crop yields by an average of 40\%, with many farmers in developing regions experiencing losses as high as 100\%~\cite{hughes2015open}. Accurate diagnosis is a prerequisite for effective treatment, yet it remains limited by two key gaps: no existing dataset is sufficiently large or diverse across crops and diseases to support diagnosis at scale, and no structured, verifiable symptom knowledge is available alongside the images.Recent advances such as InsectNet~\cite{chiranjeevi2025insectnet} and WeedNet~\cite{shen2025weednet} have shown what is possible at scale for other biotic stresses, motivating an analogous automated pipeline for plant disease diagnosis.

Existing plant disease image datasets have advanced the field but remain limited in scope. PlantVillage~\cite{hughes2015open} covers common crops under controlled conditions; PlantDoc~\cite{singh2020plantdoc} adds real-world images but only for 13 species; LeafNet~\cite{quoc2026leafnet} and CDDM~\cite{liu2024multimodal} bring scale and multimodal annotations but focus predominantly on leaf images. No existing dataset combines broad crop coverage, multi-organ imagery (leaf, stem, root, seed, ear, head), and structured symptom metadata with source provenance. This gap limits both the training of generalizable models and the development of diagnostic systems that can explain their reasoning in terms of verifiable disease knowledge.

On the methods side, deep learning has achieved high accuracy on controlled benchmarks~\cite{ghosal2018explainable}, and vision-language models have shown promise for few-shot disease recognition~\cite{arshad2025leveraging, wei2024benchmarking}. Recent work has begun to explore chain-of-thought reasoning~\cite{pan2025chatleafdisease, zhang2024visual} and agentic architectures~\cite{qin2025pdd} for plant disease diagnosis. However, these approaches either operate on narrow crop sets, require fine-tuning, or lack the ability to produce transparent, step-by-step diagnostic reasoning grounded in traceable knowledge sources.

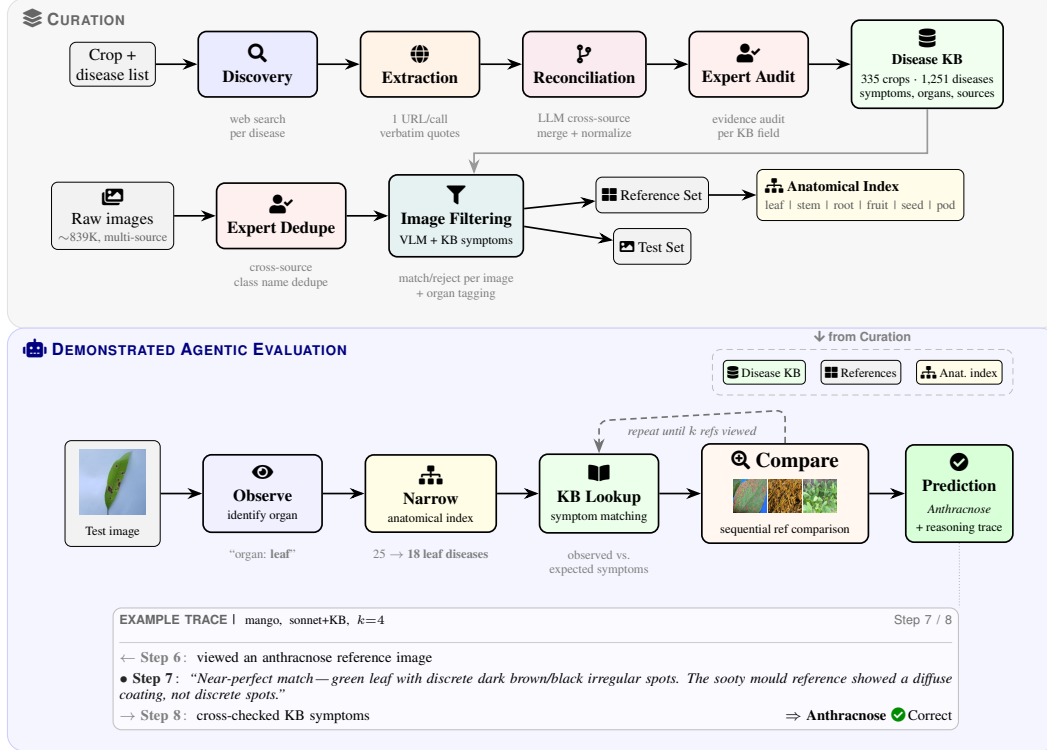
\begin{figure}[!t]
\centering
\resizebox{\columnwidth}{!}{%
\begin{tikzpicture}[
    node distance=0.4cm and 0.6cm,
    stage/.style={draw, rounded corners=4pt, minimum height=1.1cm, minimum width=2.0cm, align=center, font=\footnotesize, inner sep=5pt, thick},
    io/.style={draw, rounded corners=3pt, minimum height=0.7cm, minimum width=1.4cm, align=center, font=\footnotesize, inner sep=3pt, fill=black!5},
    smallio/.style={draw, rounded corners=3pt, minimum height=0.6cm, align=center, font=\scriptsize, inner sep=3pt, fill=black!5},
    detail/.style={font=\tiny, text=black!50, align=center},
    arr/.style={-{Stealth[length=2.5mm]}, thick},
    grouplabel/.style={font=\footnotesize\bfseries\sffamily, text=black!60},
    inputtag/.style={draw, rounded corners=2pt, fill=white, font=\tiny, inner sep=2pt, minimum height=0.4cm},
]

\node[io] (input) {Crop +\\disease list};
\node[stage, right=0.7cm of input, fill=blue!8] (disc) {{\small\faSearch}\\[1pt]\textbf{Discovery}};
\node[detail, below=0.15cm of disc] (disc_d) {web search\\per disease};
\node[stage, right=0.7cm of disc, fill=orange!10] (ext) {{\small\faGlobe}\\[1pt]\textbf{Extraction}};
\node[detail, below=0.15cm of ext] (ext_d) {1 URL/call\\verbatim quotes};
\node[stage, right=0.7cm of ext, fill=purple!8] (rec) {{\small\faCodeBranch}\\[1pt]\textbf{Reconciliation}};
\node[detail, below=0.15cm of rec] (rec_d) {LLM cross-source\\merge + normalize};

\node[stage, right=0.7cm of rec, fill=red!6] (expert_kb) {{\small\faUserCheck}\\[1pt]\textbf{Expert Audit}};
\node[detail, below=0.15cm of expert_kb] (expert_kb_d) {evidence audit\\per KB field};

\node[draw, rounded corners=4pt, right=0.7cm of expert_kb, fill=green!6, minimum width=2.4cm, minimum height=1.1cm, align=center, font=\scriptsize, inner sep=4pt, thick] (kb) {
    {\small\faDatabase}\\[1pt]
    \textbf{Disease KB}\\[1pt]
    {\tiny 335 crops $\cdot$ 1{,}251 diseases}\\[-1pt]
    {\tiny symptoms, organs, sources}
};

\draw[arr] (input) -- (disc);
\draw[arr] (disc) -- (ext);
\draw[arr] (ext) -- (rec);
\draw[arr] (rec) -- (expert_kb);
\draw[arr] (expert_kb) -- (kb);

\node[io, below=1.6cm of input, align=center] (rawimg) {{\small\faImages}\\Raw images\\[-1pt]{\tiny\color{black!60}$\sim$839K, multi-source}};

\node[stage, right=0.7cm of rawimg, fill=red!6] (expert_img) {{\small\faUserCheck}\\[1pt]\textbf{Expert Dedupe}};
\node[detail, below=0.15cm of expert_img] (expert_img_d) {cross-source\\class name dedupe};

\node[stage, right=0.7cm of expert_img, fill=teal!10] (filter) {{\small\faFilter}\\[1pt]\textbf{Image Filtering}\\{\tiny VLM + KB symptoms}};
\node[detail, below=0.15cm of filter] (filter_d) {match/reject per image\\+ organ tagging};

\node[smallio, right=1.2cm of filter, yshift=0.35cm] (refs) {{\scriptsize\faThLarge}~Reference Set};
\node[smallio, below=0.25cm of refs] (testset) {{\scriptsize\faImage}~Test Set};
\node[draw, rounded corners=3pt, fill=yellow!10, right=0.8cm of refs, align=left, font=\scriptsize, inner sep=4pt] (anatidx) {%
    {\scriptsize\faSitemap}~\textbf{Anatomical Index}\\[1pt]
    {\tiny\color{black!65} leaf \,\textbar\, stem \,\textbar\, root \,\textbar\, fruit \,\textbar\, seed \,\textbar\, pod}%
};

\draw[arr] (rawimg) -- (expert_img);
\draw[arr] (expert_img) -- (filter);
\draw[arr] (filter) -- (refs);
\draw[arr] (filter) -- (testset);
\draw[arr] (refs) -- (anatidx);
\coordinate (kb_route) at ($(filter.north)+(0.3,0.35)$);
\draw[arr, black!40] (kb.south) |- (kb_route) -- ($(filter.north)+(0.3,0)$);



\node[draw, rounded corners=3pt, below=3.2cm of rawimg, fill=black!5, minimum width=1.6cm, align=center, font=\scriptsize, inner sep=4pt] (agentinputs) {
    \includegraphics[width=1.1cm, height=1.1cm]{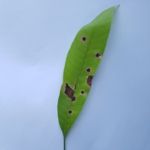}\\[2pt]
    {\tiny Test image}
};

\node[stage, right=0.7cm of agentinputs, fill=blue!6, minimum width=2.0cm] (observe) {
    {\small\faEye}\\[1pt]
    \textbf{Observe}\\[-1pt]
    {\tiny identify organ}
};
\node[detail, below=0.15cm of observe] (obs_d) {``organ: \textbf{leaf}''};

\node[stage, right=0.7cm of observe, fill=yellow!12, minimum width=2.2cm] (narrow) {
    {\small\faSitemap}\\[1pt]
    \textbf{Narrow}\\[-1pt]
    {\tiny anatomical index}
};
\node[detail, below=0.15cm of narrow] (narrow_d) {25 $\rightarrow$ \textbf{18 leaf diseases}};

\node[stage, right=0.7cm of narrow, fill=green!8, minimum width=2.0cm] (kbreason) {
    {\small\faBookOpen}\\[1pt]
    \textbf{KB Lookup}\\[-1pt]
    {\tiny symptom matching}
};
\node[detail, below=0.15cm of kbreason] (kbr_d) {observed vs.\\expected symptoms};

\node[draw, rounded corners=4pt, right=0.7cm of kbreason, fill=orange!6, minimum width=2.8cm, minimum height=1.4cm, align=center, thick] (compare) {
    \textbf{{\small\faSearchPlus}~Compare}\\[3pt]
    \includegraphics[width=0.55cm, height=0.55cm]{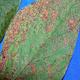}\hspace{1pt}%
    \includegraphics[width=0.55cm, height=0.55cm]{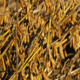}\hspace{1pt}%
    \includegraphics[width=0.55cm, height=0.55cm]{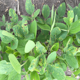}\\[-1pt]
    {\tiny sequential ref comparison}
};

\node[draw, rounded corners=4pt, right=0.6cm of compare, fill=green!15, minimum width=1.8cm, minimum height=1.2cm, align=center, font=\footnotesize, inner sep=4pt, thick] (pred) {
    {\small\faCheckCircle}\\[1pt]
    \textbf{Prediction}\\[1pt]
    {\tiny\textit{Anthracnose}}\\[-1pt]
    {\tiny + reasoning trace}
};

\node[inputtag, fill=yellow!10, above=1.0cm of pred] (intag_anat) {{\tiny\faSitemap}~Anat.\ index};
\node[inputtag, fill=black!5, left=0.25cm of intag_anat] (intag_refs) {{\tiny\faThLarge}~References};
\node[inputtag, fill=green!8, left=0.25cm of intag_refs] (intag_kb) {{\tiny\faDatabase}~Disease KB};
\node[draw=black!25, rounded corners=4pt, inner sep=5pt, densely dashed,
      fit=(intag_kb)(intag_refs)(intag_anat),
      label={[font=\tiny\bfseries\sffamily, text=black!50, anchor=south]above:{\faArrowDown} from Curation}] (inputsbox) {};

\draw[arr] (agentinputs) -- (observe);
\draw[arr] (observe) -- (narrow);
\draw[arr] (narrow) -- (kbreason);
\draw[arr] (kbreason) -- (compare);
\draw[arr] (compare) -- (pred);

\coordinate (loopL) at ($(compare.north)+(0,0.55)$);
\coordinate (loopR) at ($(kbreason.north)+(0,0.55)$);
\draw[arr, densely dashed, black!55, rounded corners=4pt]
    (compare.north) -- (loopL) -- (loopR) -- (kbreason.north);
\node[font=\tiny\itshape, text=black!60, anchor=north]
    at ($(loopL)!0.5!(loopR)+(0,-0.05)$) {repeat until $k$ refs viewed};

\begin{scope}[on background layer]
    \node[draw=black!30, dashed, rounded corners=6pt, inner sep=10pt,
          fit=(observe)(narrow)(kbreason)(compare)(obs_d)(narrow_d)(kbr_d),
          label={[font=\tiny\itshape, text=black!45]above:open-ended reasoning loop (bounded by reference budget $k$)}] (reasonbox) {};
\end{scope}

\node[draw=black!25, fill=blue!2, rounded corners=4pt,
      below=0.05cm of reasonbox.south, inner sep=3pt,
      font=\scriptsize, align=left, text width=14.0cm] (tracestrip) {%
    {\tiny\bfseries\sffamily\color{black!60} EXAMPLE TRACE}\,\,\textbar\,\,
    {\tiny mango,\, sonnet+KB,\, $k{=}4$}%
    \hfill{\tiny\sffamily\color{black!55} Step 7 / 8}\\[1pt]
    {\color{black!20}\hrulefill}\\[1pt]
    {\scriptsize\color{black!50}$\leftarrow$ \textbf{Step 6}\,:}\, viewed an anthracnose reference image\\[2pt]
    {\scriptsize\color{black!85}$\bullet$ \textbf{Step 7}\,:}\, \textit{``Near-perfect match\,---\,green leaf with discrete dark brown/black irregular spots. The sooty mould reference showed a diffuse coating, not discrete spots.''}\\[2pt]
    {\scriptsize\color{black!50}$\rightarrow$ \textbf{Step 8}\,:}\, cross-checked KB symptoms\hfill$\Rightarrow$ \textbf{Anthracnose} \textcolor{green!55!black}{\faCheckCircle}\,Correct%
};
\draw[densely dotted, black!35] (pred.south) -- ($(tracestrip.north -| pred.south)$);

\begin{scope}[on background layer]
    \node[draw=none, fill=none, inner sep=10pt,
          fit=(input)(kb)(rawimg)(anatidx)(filter)(disc_d)(ext_d)(rec_d)(expert_kb)(expert_kb_d)(expert_img)(expert_img_d)(filter_d)] (prepgroup_nat) {};
    \node[draw=none, fill=none, inner sep=10pt,
          fit=(agentinputs)(pred)(reasonbox)(obs_d)(narrow_d)(kbr_d)(inputsbox)(tracestrip)] (infgroup_nat) {};
\end{scope}

\path let \p1=(prepgroup_nat.west), \p2=(infgroup_nat.west) in
      coordinate (sl) at ({min(\x1,\x2)}, 0);
\path let \p1=(prepgroup_nat.east), \p2=(infgroup_nat.east) in
      coordinate (sr) at ({max(\x1,\x2)}, 0);

\begin{scope}[on background layer]
    \node[draw=none, fill=none, inner sep=0pt] at (sl |- prepgroup_nat.center) (cur_l) {};
    \node[draw=none, fill=none, inner sep=0pt] at (sr |- prepgroup_nat.center) (cur_r) {};
    \node[draw=none, fill=none, inner sep=0pt] at (sl |- infgroup_nat.center) (inf_l) {};
    \node[draw=none, fill=none, inner sep=0pt] at (sr |- infgroup_nat.center) (inf_r) {};

    \node[draw=black!12, fill=black!3, rounded corners=8pt, inner sep=10pt,
          fit=(input)(kb)(rawimg)(anatidx)(filter)(disc_d)(ext_d)(rec_d)(expert_kb)(expert_kb_d)(expert_img)(expert_img_d)(filter_d)(cur_l)(cur_r)] (prepgroup) {};
    \node[grouplabel, anchor=north west] at ($(prepgroup.north west)+(0.15,-0.05)$) {{\small\faLayerGroup}~\textsc{Curation}};

    \node[draw=blue!20, fill=blue!4, rounded corners=8pt, inner sep=10pt,
          fit=(agentinputs)(pred)(reasonbox)(obs_d)(narrow_d)(kbr_d)(inputsbox)(tracestrip)(inf_l)(inf_r)] (infgroup) {};
    \node[grouplabel, anchor=north west, text=blue!50!black] at ($(infgroup.north west)+(0.15,-0.05)$) {{\small\faRobot}~\textsc{Demonstrated Agentic Evaluation}};
\end{scope}

\end{tikzpicture}
}%
\caption{System overview. \textbf{Curation} (top): web pages become a source-cited KB (335 crops, 1{,}251 diseases) with an expert audit; raw images are deduped, filtered against the KB, and split into reference/test sets with an anatomical index. \textbf{Demonstrated agentic evaluation} (bottom): the agent observes the organ, narrows candidates, consults KB symptoms, and sequentially compares references, producing a prediction with a step-by-step trace.}
\label{fig:pipeline}
\end{figure}

We address these gaps together. We compile one of the largest plant disease image datasets, complement it with automatically generated symptom descriptions that trace back to their sources, and show how this combination enables a training-free, explainable agentic reasoning system to diagnose diseases across crops (Figure~\ref{fig:pipeline}). Specifically, we present:
\begin{enumerate}[leftmargin=*]
  \item A multi-crop image dataset spanning 335 crops and 1{,}251 disease classes ($\sim$839K images), assembled from established benchmarks, expert-curated collections, and community sources, with multi-organ coverage.
  \item A source-first disease registry pipeline that, given a crop name, automatically produces structured symptom knowledge with per-field provenance, where every fact traces back to a specific web source with a verbatim supporting quote.
  \item A training-free agentic diagnostic pipeline where each prediction is made by an autonomous reasoning agent that produces an explainable, human-readable reasoning trace showing which references were examined and why.
  \item A systematic evaluation across four crops of varying difficulty, multiple reference budgets, knowledge base sources, and model tiers.
\end{enumerate}

\section{Related Work}
\label{sec:related}

The automatic identification of plant diseases from images has been studied extensively using deep learning. Convolutional neural networks trained on labeled image datasets have achieved high accuracy on controlled benchmarks, with architectures such as ResNet, VGG, and DenseNet commonly applied to leaf disease classification~\cite{ghosal2018explainable}. However, these models require large labeled training sets for each new crop or disease and produce opaque predictions: a user receives a label but no explanation of what visual features led to it.

The data landscape has evolved in parallel. PlantVillage~\cite{hughes2015open} established the first large-scale open repository of plant disease images, covering 38 classes across common crops. PlantDoc~\cite{singh2020plantdoc} extended this with real-world field images, while PlantWild~\cite{wei2024benchmarking} introduced in-the-wild images paired with text descriptions for multimodal recognition. More recently, LeafNet~\cite{quoc2026leafnet} provides 186K images across 97 classes with a vision-language benchmark, and CDDM~\cite{liu2024multimodal} offers 137K images with 1M question-answer pairs for conversational diagnosis. AgMMU~\cite{gauba2025agmmu} takes a different approach, constructing a benchmark from real extension expert dialogues that spans identification, symptom description, and management across multiple crops. Despite this progress, existing datasets are predominantly leaf-only, cover limited crop diversity, and lack structured symptom metadata with source provenance, the kind of knowledge a pathologist would consult when making a diagnosis.

Vision-language models have brought new capabilities to this space. General-purpose VLMs can perform few-shot and zero-shot disease recognition when provided with example images or textual descriptions~\cite{arshad2025leveraging}. Domain-specific models such as AgroGPT~\cite{awais2025agrogpt} and Agri-LLaVA~\cite{wang2024agri} have been fine-tuned on agricultural data to improve conversational diagnosis. AgReason~\cite{zaremehrjerdi2025towards} introduces a reasoning benchmark for agricultural decision-making and shows that large reasoning models outperform conventional LLMs on domain-specific inference. These works demonstrate that VLMs have substantial knowledge of plant diseases, but most operate in a single-pass mode: the model sees an image and produces an answer in one step.

A growing line of work moves beyond single-pass classification toward structured reasoning. ChatLeafDisease~\cite{pan2025chatleafdisease} uses chain-of-thought prompting with a disease description database to guide GPT-4o through scoring-based classification, achieving 88.9\% accuracy on tomato diseases without any training. WDLM~\cite{zhang2024visual} integrates a five-step reasoning chain for wheat disease diagnosis, generating explanations alongside predictions. PDD-AGENT~\cite{qin2025pdd} proposes a multimodal agent system that orchestrates specialized diagnostic tools through a decision-making module. Agri-CPJ~\cite{zhang2026agri} introduces a training-free caption-prompt-judge pipeline with an LLM-as-a-judge for explainable agricultural diagnosis; methods of this kind benefit directly from broader source-grounded corpora and demonstrated agentic evaluation
These approaches share our motivation of making the diagnostic process transparent and grounded in domain knowledge, but each targets a narrow crop set and does not combine autonomous reference selection with traceable symptom knowledge at scale.

Our work brings together these threads. The core bottleneck, the lack of a large-scale dataset with both images and structured symptom knowledge, limits all of the above approaches. We address this directly with an 839K-image dataset for 1{,}251 diseases across 335 crops paired with source-cited symptom descriptions, and an agentic diagnostic system that uses both. The agent autonomously selects reference images, reasons in natural language, and grounds its decisions in verifiable knowledge without any task-specific training. 

\section{Dataset}
\label{sec:dataset}

\subsection{Image Sources}

\begin{figure}[t]
\centering
\begin{minipage}[c]{0.68\columnwidth}
\centering
\includegraphics[width=\textwidth]{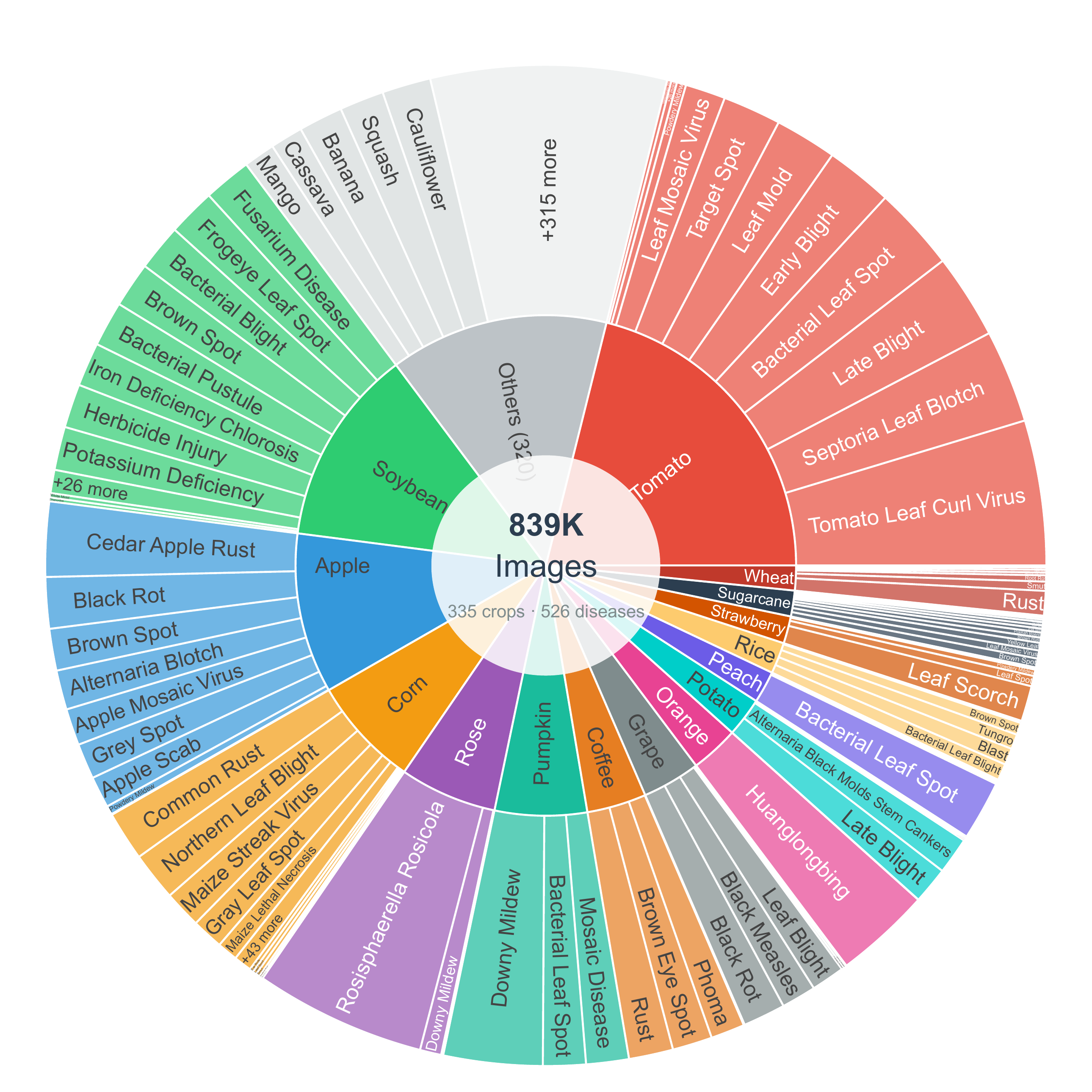}
\end{minipage}%
\hfill
\begin{minipage}[c]{0.30\columnwidth}
\begin{tcolorbox}[
    colback=black!4, colframe=black!30, boxrule=0.3pt,
    left=3pt, right=3pt, top=2pt, bottom=2pt,
]
\scriptsize
\textbf{Standard dataset}\\[1pt]
\textcolor{black!50}{Image + Disease label}
\end{tcolorbox}
\vspace{2pt}
\centering{\scriptsize\textbf{vs.}}
\vspace{2pt}
\begin{tcolorbox}[
    colback=green!4, colframe=green!50!black, boxrule=0.3pt,
    left=3pt, right=3pt, top=2pt, bottom=2pt,
]
\scriptsize
\textbf{Ours}\\[1pt]
Image + Disease label\\
+ \textbf{Organ tags} {\tiny(leaf, stem, root, \ldots)}\\
+ \textbf{Symptom descriptions}\\
+ \textbf{Source URL} + \textbf{Verbatim quote}
\end{tcolorbox}
\vspace{4pt}
\begin{tcolorbox}[
    colback=blue!3, colframe=blue!40, boxrule=0.4pt,
    title={\scriptsize \textbf{Example Entry}},
    fonttitle=\scriptsize, left=3pt, right=3pt, top=2pt, bottom=2pt,
]
\scriptsize
\textbf{Disease:} Tomato Early Blight\\[1pt]
\textbf{Pathogen:} \textit{A.\ linariae}\\[1pt]
\textbf{Organs:} leaf, stem\\[1pt]
\textbf{Symptoms:} Circular brown spots with concentric rings, target-like appearance, yellow halo\\[1pt]
\textbf{Source:} \textcolor{blue!60}{hort.extension.wisc.edu}\\[1pt]
\textbf{Quote:} \textit{``roughly circular, brown spots appear on leaves and stems\ldots concentric rings appear giving a target-like appearance.''}
\end{tcolorbox}
\end{minipage}
\caption{(Left) Image distribution across 335 crops and 1{,}251 disease classes ($\sim$839K images). (Right) Each disease carries source-cited metadata (organ tags, symptoms, verbatim quotes), unlike standard image+label datasets.}
\label{fig:sunburst}
\end{figure}

No single existing plant disease dataset covers enough crops and diseases for a general-purpose diagnostic system. We assembled our dataset from four complementary source categories to maximize coverage (Figure~\ref{fig:sunburst}).

We start with \textbf{established benchmarks}: PlantVillage~\cite{hughes2015open}, the most widely used plant disease image repository covering common crops under controlled imaging conditions, and PlantDoc~\cite{singh2020plantdoc}, which extends this with real-world field images. We supplement these with \textbf{recent large-scale datasets}: LeafNet~\cite{quoc2026leafnet}, a collection designed for vision-language understanding of plant diseases, and PlantWild~\cite{wei2024benchmarking}, which focuses on in-the-wild disease recognition under uncontrolled conditions.

For agriculturally important crops, we include \textbf{expert-curated data} with multi-organ coverage (leaf, stem, root, seed, ear, and head), going beyond the leaf-only focus of most public datasets. The soybean stress identification dataset~\cite{ghosal2018explainable} contributes images with expert-verified disease labels, and the mango evaluation subset draws from MangoLeafBD~\cite{ahmed2023mangoleafbd}. For crops not well-represented in the above sources, we incorporate \textbf{community-contributed datasets} covering rice~\cite{ritharson2021sbrd}, wheat rust~\cite{hayit2020yellowrust}, banana~\cite{zarita2022banana}, bean~\cite{marquis2023bean}, lettuce~\cite{jstar2023lettuce}, cucumber~\cite{kareem2023cucumber}, durian~\cite{cthng2023durian}, eggplant~\cite{kamalmoha2023eggplant}, cotton~\cite{basak2023cotton}, pumpkin~\cite{basak2023pumpkin}, rose~\cite{basak2023rose}, strawberry~\cite{afzaal2023strawberry}, sugarcane~\cite{puspasari2023sugarleafidn}, chickpea~\cite{hayit2022fusarium}, cauliflower~\cite{basak2024cauliflower}, coconut~\cite{basak2024coconut}, and vanilla~\cite{permana2024vanilla}. We further include a curated cucumber set~\cite{cucumberzenodo2025}, an augmented PlantVillage derivative~\cite{vipoooool2018newplant}, six HuggingFace plant-disease aggregator datasets~\cite{rady2024plantimagetext, a2h0h0r1plantdiseasenew, avinashhm2024plantcomplete, sakethdevx2024plant, raghavendrad60vqaplant, saon110bdcrop}, and per-image attributed photography from the Bugwood Image Database~\cite{bugwood}. Per-dataset licenses and citation requirements are listed in Appendix~\ref{sec:appendix_licenses}.

The combined dataset spans 335 crops and 1{,}251 disease classes, totaling 838{,}936 images. For the agentic evaluation in this work, we select four crops that represent different scales of difficulty: Soybean (25 disease classes), Corn (30 classes), Tomato (20 classes), and Mango (4 classes). Evaluation on additional crops is planned but limited by the compute cost of agentic inference. Table~\ref{tab:dataset_comparison} positions SAGE against existing plant disease datasets and benchmarks. Several related works are mentioned above but excluded from the dataset comparison because they focus on different artifacts. AgMMU~\cite{gauba2025agmmu} is a VLM benchmark of $\sim$1.5K MCQ/open-ended questions distilled from 116K real grower-Extension dialogues, covering insect, species, disease, and management QA rather than a labeled image corpus. AgroBench~\cite{shinoda2025agrobench} similarly evaluates VLMs across seven agricultural topics with expert-annotated categories spanning 203 crops and 682 diseases. Agri-LLaVA~\cite{wang2024agri} provides a 400K-entry instruction-following dataset across 221 pest/disease classes paired with a knowledge-infused conversational model, and AgroGPT~\cite{awais2025agrogpt} synthesizes 70K instruction-tuning examples from class-only vision data to train an expert-tuned LMM. None release a per-image labeled corpus directly comparable to SAGE.

\input{table_datasets_v2}

\subsection{Splits and Image Filtering}
Because images are aggregated from many sources, the same disease often appears under several different class names (e.g., ``Leaf\_Rust'', ``leaf rust'', and ``rust\_disease'' across three sources). Before any automated filtering, domain experts have reviewed the union of class labels across sources and merge variants that refer to the same disease, producing a single canonical class per disease. After this dedupe step, raw image collections still contain noise: mislabeled images, images showing healthy tissue, or images where the disease presentation does not match the documented symptoms. We filter images using the disease registry as ground truth. For each image in a class, a vision-language model receives the image alongside the knowledge base (KB) symptom description for that class and determines whether the image is consistent with the description. Matched images are kept; mismatches are rejected.

During filtering, each matched image is also tagged with its \emph{anatomical context}: the primary plant organ visible (leaf, stem, root, seed, pod, or whole plant). This produces an anatomical index that maps each organ to the set of diseases that affect it, enabling the agent to narrow candidates at inference based on the organ visible in the test image. Figure~\ref{fig:pipeline} (Curation, top row) illustrates this filtering and tagging process.

Matched images are split into reference and test sets. Reference images are organized by class and organ; test images are held out for evaluation. For the four evaluation crops, this yields Soybean (25 classes, 74 test images), Corn (30 classes, 88 test images), Tomato (20 classes, 88 test images), and Mango (4 classes, 40 test images).

\section{Disease Registry Pipeline}
\label{sec:registry}

The diagnostic agent (Section~\ref{sec:agent}) needs structured disease knowledge to reason over images at inference. Querying a language model directly for disease symptoms risks hallucination and produces facts that cannot be verified. Our pipeline instead follows a source-first principle: it fetches real documents from the web, extracts only what is explicitly stated in those documents, and traces every fact back to its source with a verbatim quote. Given a crop name, three automated stages followed by an expert evidence audit (Figure~\ref{fig:pipeline}) produce a fully cited disease registry.

The pipeline begins by issuing targeted web search queries for each disease in the crop, using the Claude CLI's web search capability to retrieve authoritative pages such as university extension factsheets, CABI datasheets, and APS disease compendia. Each retrieved page is then passed to the language model with an extraction prompt that requires the model to identify all diseases discussed on the page and extract structured fields: pathogen name, disease type, affected plant organs, and visual symptom descriptions. The model is required to provide a verbatim quote from the source text supporting each extracted field and is explicitly prevented from filling in information from its own knowledge. The raw extractions from multiple sources are then reconciled into a canonical registry: naming inconsistencies are resolved through LLM-based matching, conflicting claims across sources are tracked, and each field in the final registry retains its provenance (schema in Appendix~\ref{sec:appendix_schema}; expert audit results in Appendix Figure~\ref{fig:field_verdicts}). Once generated for a crop, the registry is reused across all evaluation runs. The pipeline takes a crop name and disease list as input and runs end-to-end without per-crop tuning; because every field is anchored to a verbatim source quote, the registry can be audited directly rather than re-checked against a model's recall. We sample-audit registries end-to-end with an expert agronomist; field-level agreement on the audited crops is reported in Appendix Figure~\ref{fig:field_verdicts}. Figure~\ref{fig:kb_sources} shows where the released registries draw from: predominantly US land-grant extension publications, complemented by international compendia (CABI, Lucid Pacific Pests, PNW Plant Disease Handbook), peer-reviewed journals, and the multi-university Crop Protection Network.

\begin{figure}[!htbp]
\centering
\includegraphics[width=\textwidth]{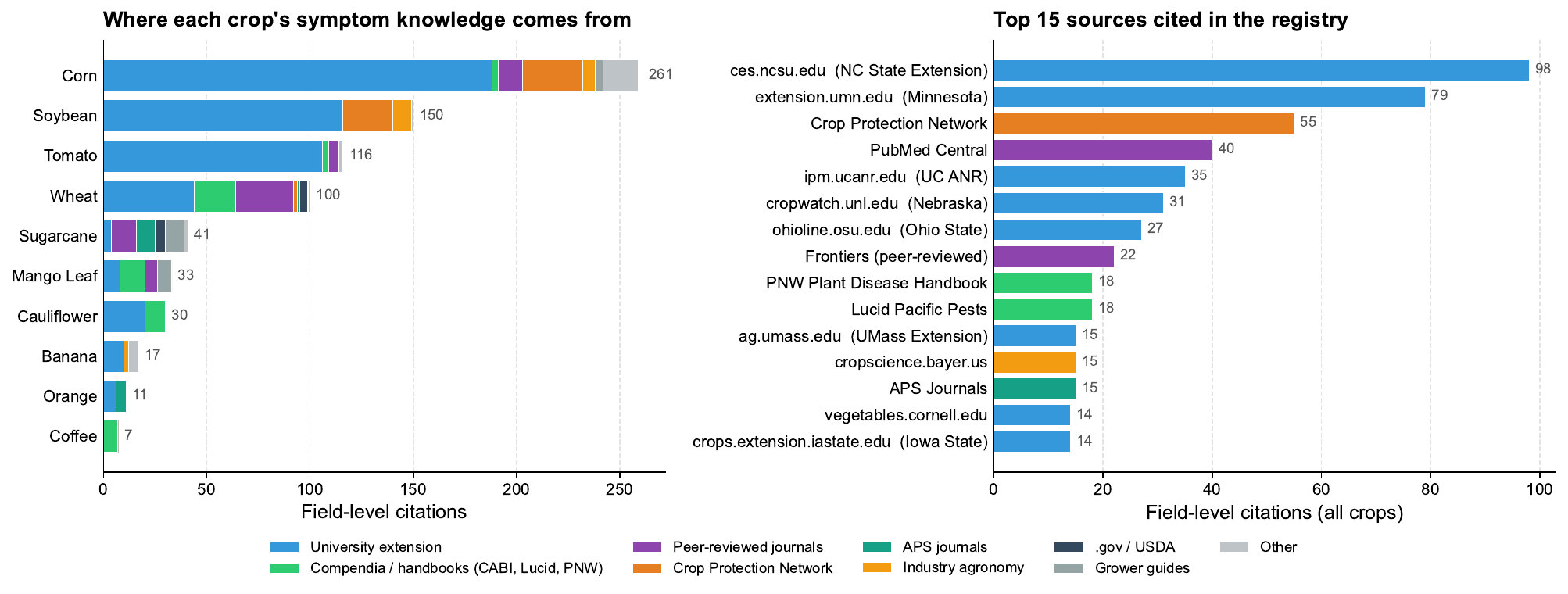}
\caption{Sources backing the disease registry across all 10 crops with KBs released to date. Left: per-crop field-level citations stacked by source category. Right: top 15 cited domains across these KBs. Full KB and code: \url{https://sage-dataset.github.io/}.}
\label{fig:kb_sources}
\end{figure}

\section{Agentic Diagnostic Pipeline}
\label{sec:agent}

Each test image is diagnosed by an autonomous reasoning agent that operates in a multi-turn loop. The agent receives three inputs: (1)~the test image to classify, (2)~individually labeled reference images organized by disease class, and (3)~optionally, the symptom descriptions from the disease registry (Section~\ref{sec:registry}) together with an anatomical index that maps plant organs to the diseases that affect them. It also receives the full list of candidate disease classes and a reference budget $k$ that limits how many reference images it may examine.

The agent follows a guided chain of thought. It first reads the test image and identifies the anatomical context (leaf, stem, root, seed) and visible symptoms: color changes, lesion shape, texture, and spatial patterns. If the knowledge base is available, the agent consults the \textbf{anatomical index} to narrow candidates to only those diseases that affect the observed organ, then reviews symptom descriptions to further prioritize. It then views reference images \textbf{one at a time}, comparing the test image against each labeled example and reasoning about what matches and what does not before selecting the next reference to examine. This sequential, deliberative comparison ensures the agent explores multiple candidates rather than confirming an initial guess. The agent concludes by producing a structured prediction with a brief explanation of its reasoning. The full system and user prompt is given in Appendix~\ref{sec:appendix_prompt}.

The full sequence of the agent's actions and reasoning is captured as a \textbf{trace}. Each trace records which reference images the agent chose to view, the text it produced between views, and its final prediction. These traces are the primary vehicle for explainability: they allow a user to see not just what the system predicted but how it arrived at that prediction, which references it found most informative, and what alternative diagnoses it considered and ruled out. This stands in contrast to single-pass classification, where the model produces a label with no visibility into the decision process. The trace below illustrates this process on a soybean test image at $k{=}8$ with internet KB. Additional traces are provided in Appendix~\ref{sec:appendix_traces}.

\input{traces/trace_intext.tex}

\section{Experiments}
\label{sec:experiments}

We evaluate the diagnostic pipeline on four crops spanning a range of class sizes: Soybean (25 classes), Corn (30 classes), Tomato (20 classes), and Mango (4 classes). We use 3 test images per class for Soybean and Corn, up to 5 per class for Tomato, and 10 per class for Mango, selected with a fixed random seed for reproducibility (per-crop counts in Appendix~\ref{sec:appendix_test_sizes}). We vary three axes: the reference budget $k$ (0, 1, 4, 8), the KB source (none or internet), and the model tier (Haiku, Sonnet, Opus). The baseline is the agent at $k{=}0$ without a knowledge base, where the model receives only the test image and the list of candidate class names with no reference images or symptom descriptions.

\input{tables/table_main_results.tex}

\begin{wrapfigure}[18]{r}{0.5\columnwidth}
\vspace{-22pt}
\centering
\includegraphics[width=0.48\columnwidth]{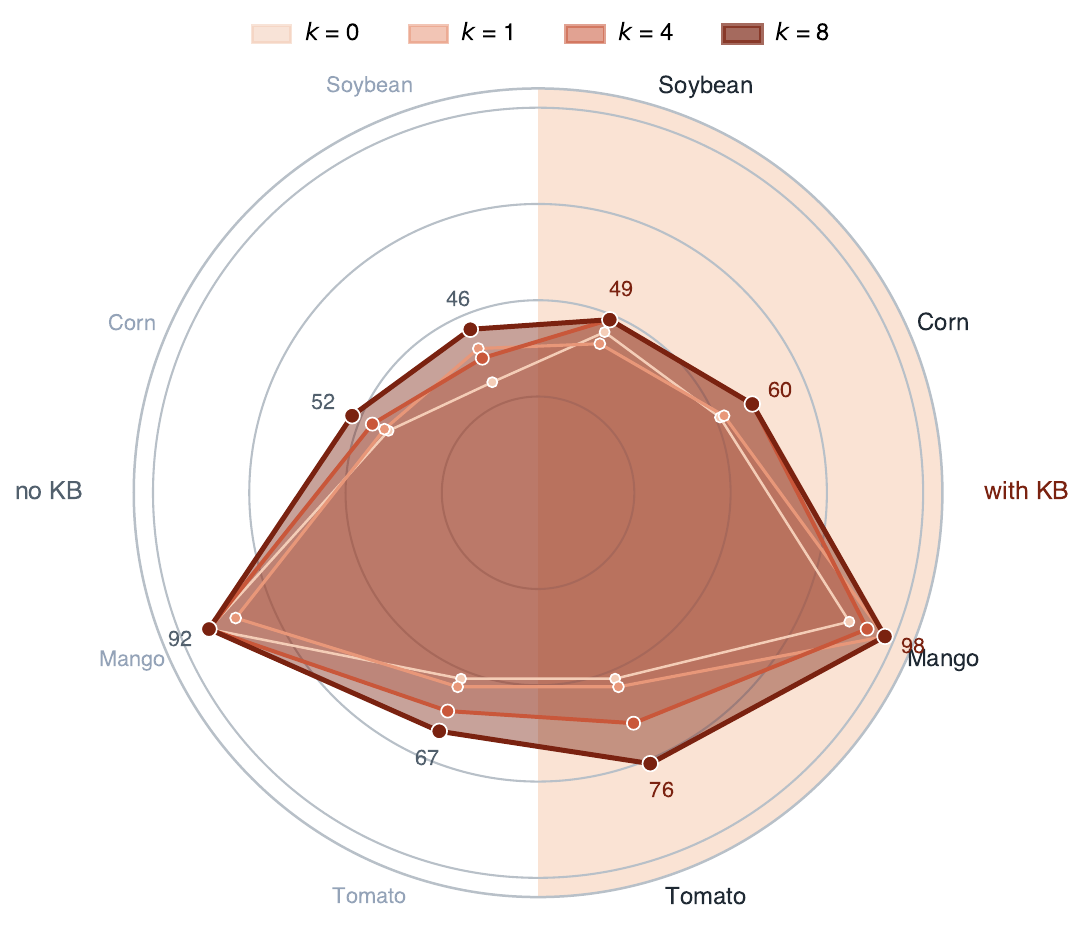}
\caption{Per-crop accuracy across reference budgets. Right semicircle = with internet KB; left = no KB. Concentric polygons (light to dark) correspond to $k=0,1,4,8$.}
\label{fig:split_radar}
\vspace{-8pt}
\end{wrapfigure}

Table~\ref{tab:main_results} presents the main results. The $k{=}0$ column isolates the KB contribution: without any reference images, the agent relies entirely on the test image and symptom descriptions. On Soybean, the KB alone adds 14.1 percentage points (31.1\% to 45.2\%), and on Corn 9.1 points (42.0\% to 51.1\%). As the reference budget increases, accuracy improves further: Tomato reaches 76.1\% and Corn 60.2\% at $k{=}8$ with KB (per-crop curves in Appendix~\ref{sec:appendix_accuracy_curves}). \emph{The full pipeline (KB, $k{=}8$) improves on the no-KB $k{=}0$ baseline by 16.2 percentage points on average across the four crops.} Figure~\ref{fig:split_radar} summarizes the per-crop KB effect across all reference budgets in one visual: the right semicircle shows accuracy with internet KB, the left shows the matched no-KB condition, and concentric polygons trace the four budgets ($k\in\{0,1,4,8\}$). Right-side bulges across Soybean, Corn, and Tomato indicate consistent KB benefit; Mango is saturated and roughly symmetric.

\leavevmode\vspace{-1.6\baselineskip}

\begin{figure}[b]
\centering
\includegraphics[width=\columnwidth]{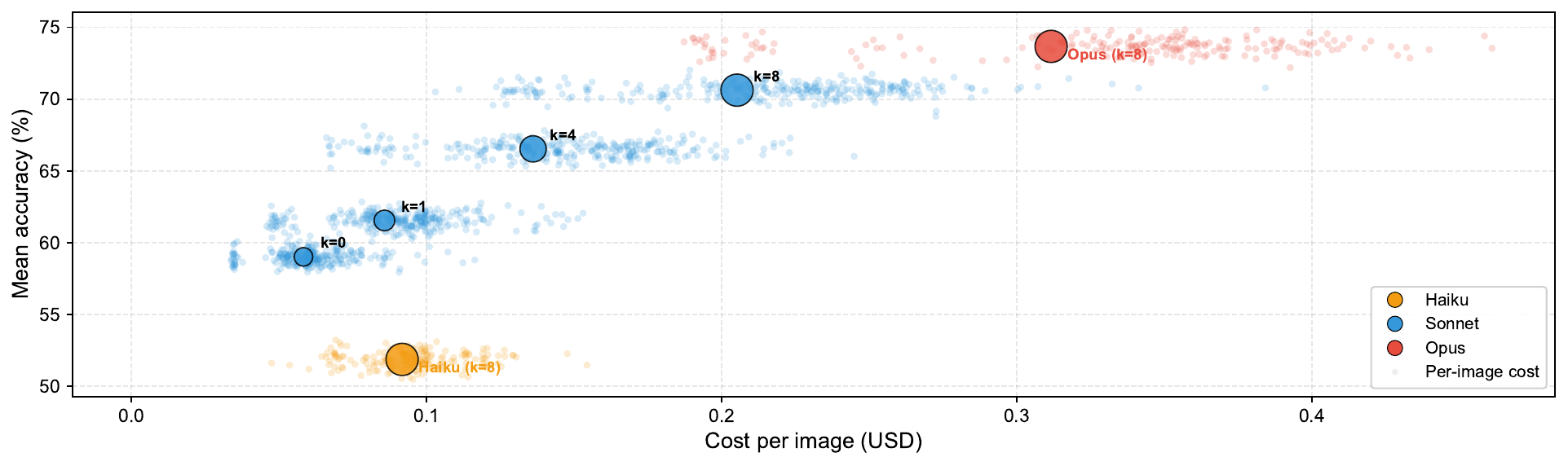}
\caption{Cost-accuracy tradeoff (mean accuracy across all four crops, internet KB). Small dots show individual per-image API costs; large bubbles show aggregate means with bubble size proportional to reference budget $k$. Haiku and Opus shown at $k{=}8$. {\footnotesize Cost estimates aggregated from Soybean, Corn, and Mango.}}
\label{fig:cost_accuracy}
\end{figure}

Figure~\ref{fig:cost_accuracy} shows the cost-accuracy tradeoff, averaging accuracy across all four crops. Small dots show API costs per-image, while large bubbles mark aggregate means at each reference budget. Increasing $k$ improves accuracy but at growing cost. Model quality provides a complementary lever: at $k{=}8$, Opus achieves higher mean accuracy than Sonnet at the same budget for comparable cost, while Haiku is the cheapest option but substantially less accurate.

\begin{figure*}[!ht]
\centering
\includegraphics[width=0.49\textwidth]{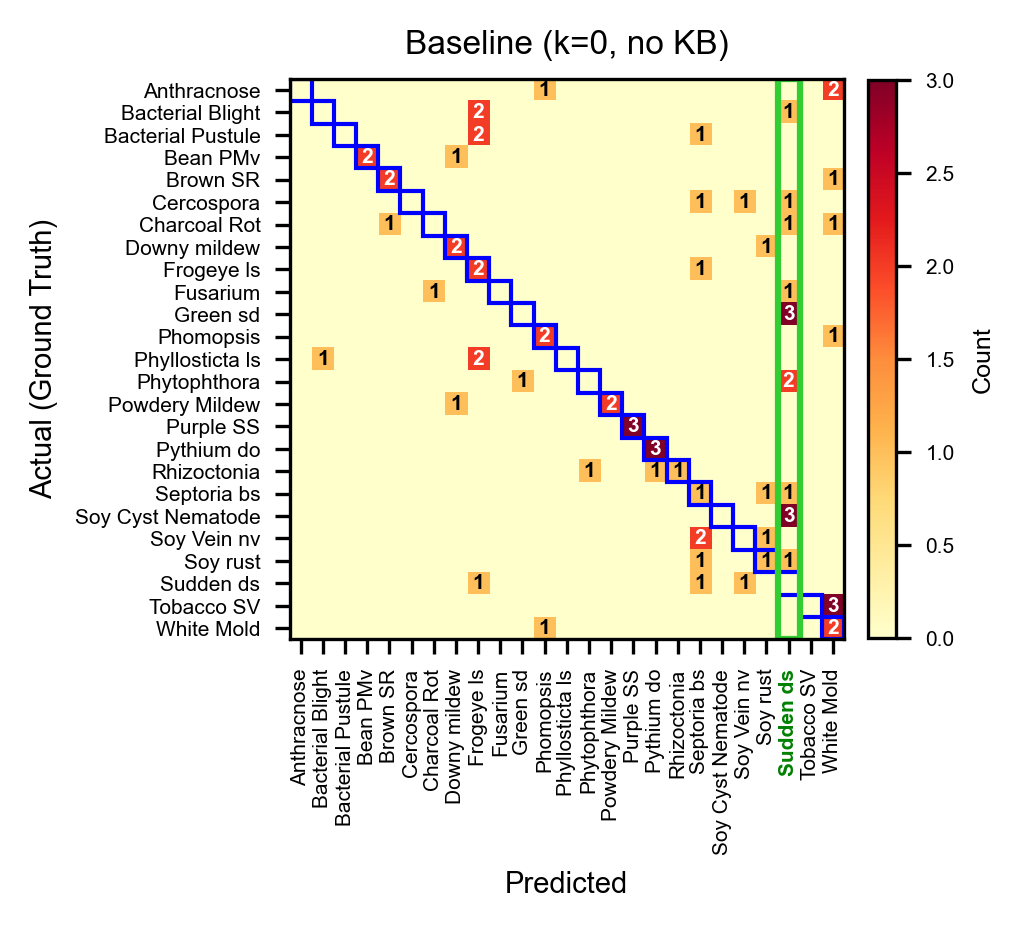}%
\hfill
\includegraphics[width=0.49\textwidth]{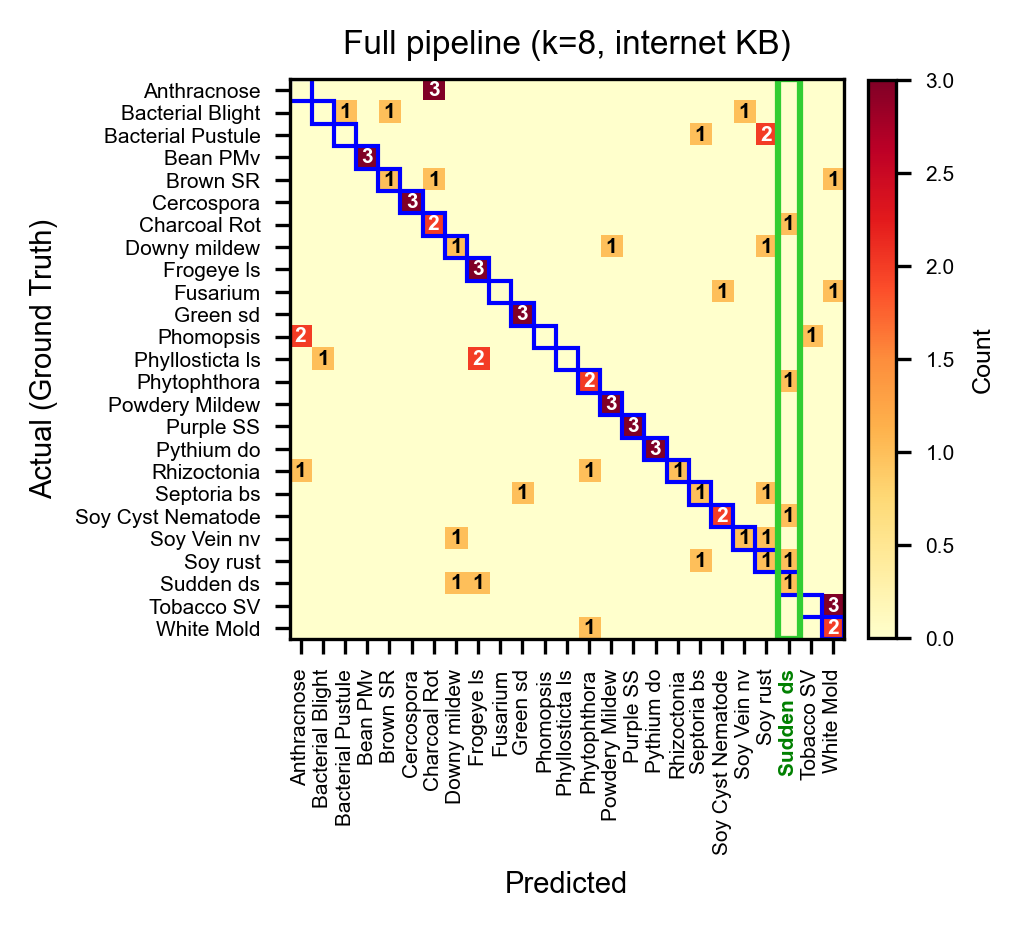}
\caption{Confusion matrices for Soybean (Sonnet, 25 classes). Left: baseline, $k{=}0$, no KB (31.1\%). Right: full pipeline, $k{=}8$, internet KB (48.6\%). The highlighted Sudden\_death\_syndrome column shows over-prediction reduced as the agent uses KB symptoms and reference comparisons to distinguish visually similar diseases.}
\label{fig:confusion_matrix}
\end{figure*}

Figure~\ref{fig:confusion_matrix} shows the confusion matrices for Soybean. In the baseline (left), Sudden\_death\_syndrome is heavily over-predicted, absorbing predictions from many other classes. With the full pipeline (right), this is reduced as the agent uses KB symptoms and reference comparisons to distinguish visually similar diseases.

\section{Limitations}
\label{sec:limitations}

The KB benefit depends on class count. On Mango (4 classes) the agent saturates near ceiling regardless of KB; on Soybean and Corn (25 and 30 classes) the KB carries most of the early-budget gain. Visual similarity alone is largely sufficient at small class counts. Per-image inference cost (Figure~\ref{fig:cost_accuracy}) is the other practical limit: Sonnet at $k{=}8$ averages \$0.21 per image and Opus roughly doubles that, which is workable for high-value uses such as extension scouting or breeding pipelines, less so for routine smartphone-scale field deployment. The registry is built from English-language web pages and extension materials, so crops documented mainly in other languages will produce a thinner KB.

We currently rely on the Claude API ecosystem because it supports programmatic agentic loops at the scale we evaluate; we expect this ecosystem to mature, and to reduce risk we test on all three model tiers (Haiku, Sonnet, Opus) so the trends are not tied to a single model. We see the dataset and registry pipeline as the durable contribution; extending evaluation (more crops, models, seeds) and agent design (cheaper inference, multi-disease handling, stage-aware symptoms) are natural next steps.

\section{Conclusion and Broader Impact}
\label{sec:discussion}

The agent improves over the baseline on all four crops. The KB provides the largest gain at low reference budgets, where symptom descriptions and the anatomical index guide the agent to relevant references first. KB alone adds 14.1 pp on Soybean and 9.1 pp on Corn at $k{=}0$. On Tomato, KB and references compound: KB alone is neutral at $k{=}0$, but at $k{=}8$ the KB-equipped agent reaches 76.1\% versus 67.0\% without KB.

Model quality is the most high-impact factor: upgrading the underlying model yields larger gains than adding a KB or increasing the reference budget. \emph{The system therefore improves automatically as foundation models improve, with no retraining required.}

Against a single-pass few-shot baseline (Appendix Table~\ref{tab:fewshot}), the agent wins at matched $k$ because the chain of thought lets it selectively compare references rather than processing all examples in one pass. Few-shot is also a black box, while every prediction here is accompanied by a reasoning trace that lists which references were viewed and why, making each diagnosis auditable end to end.

The dataset, the source-cited symptom registry, and the training-free agentic loop together show that scalable plant disease diagnosis can be done without per-crop supervised training, and with full visibility into why each label was chosen.

\clearpage
\bibliographystyle{unsrtnat}
\bibliography{referencesUpdated}

\clearpage
\appendix

\section{Evaluation Test-Set Sizes}
\label{sec:appendix_test_sizes}

Section~\ref{sec:experiments} states the target per-class counts; the table below adds the realized post-filter range. Some classes ended below the target because KB-guided filtering rejected images whose visual content did not match the documented symptoms.

\begin{table}[h]
\centering
\small
\begin{tabular}{lrrr}
\toprule
Crop & Classes & Total test images & Per class \\
\midrule
Soybean & 25 & 74 & 2--3 (avg 3.0) \\
Corn    & 30 & 88 & 2--3 (avg 2.9) \\
Tomato  & 20 & 88 & 3--5 (avg 4.6) \\
Mango   &  4 & 40 & 10            \\
\bottomrule
\end{tabular}
\caption{Number of test images evaluated per class for each crop. Some classes had fewer than the target after KB-guided filtering rejected off-target images, hence the ranges.}
\label{tab:test_sizes}
\end{table}

\section{Dataset Licenses}
\label{sec:appendix_licenses}

The SAGE collection is built from sources covering a mix of permissive and restricted licenses. The majority allow redistribution under attribution (CC BY 4.0, CC BY-SA 3.0, MIT, Apache 2.0); a smaller set carries non-commercial restrictions (CC BY-NC 3.0, CC BY-NC-ND 4.0) which we honor by linking to original hosts rather than mirroring. Bugwood imagery uses per-image attribution preserved in our metadata, and the World Bank rice subset follows the World Bank Dataset Terms of Use.

\input{dataset_licenses_snippet}

\section{Registry Schema}
\label{sec:appendix_schema}

Every entry in the released registry conforms to the schema below. Each non-identifier field is stored as a triple of \texttt{\{value, source\_url, verbatim\_quote\}} so that any claim can be traced back to a specific passage on a specific page. This is what makes the expert audit step (Section~\ref{sec:registry}) tractable: a reviewer reads the quote next to the claim and judges whether the source actually supports it. It also lets downstream users and reviewers re-verify any field at any time without re-running the pipeline.

\begin{table}[h]
\centering
\begin{tabular}{ll}
\toprule
Field & Example \\
\midrule
Crop (host) & Soybean \\
Disease & White mold (Sclerotinia Stem Rot) \\
Pathogen & \textit{Sclerotinia sclerotiorum} \\
Pathogen type & Fungal \\
Affected organs & Stem, seed \\
Visual symptoms & lesion shape, color, texture, \ldots \\
\bottomrule
\end{tabular}
\caption{Registry schema. Each field carries per-field provenance: the extracted value, the source URL, and a verbatim supporting quote.}
\label{tab:schema}
\end{table}

An agronomist audited the registry field by field for five crops, marking each claim as agreement, disagreement, or unsure. Agreement is highest on host-organ and pathogen-type claims, which are factual and easy to verify; the bulk of disagreement concentrates in fine-grained visual symptom claims where source pages either conflate disease stages or describe symptoms in ways that do not transfer to a single image. The 70--90\% agreement range across crops is the headline evidence that the source-grounded extraction is faithful enough to ship to the agent without manual rewriting.

\begin{figure}[h]
\centering
\includegraphics[width=\linewidth]{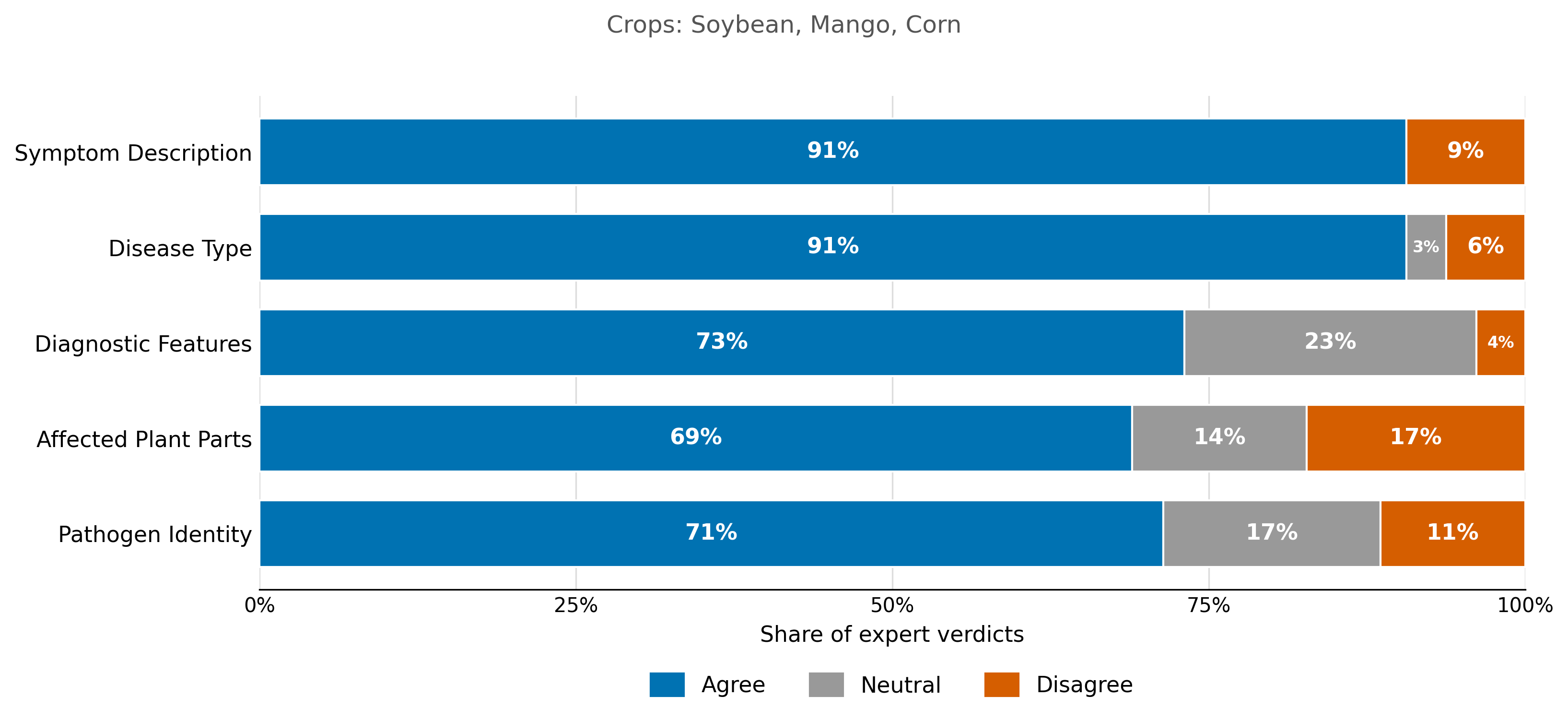}
\caption{Expert agronomist verdicts on KB-sourced claims across five crops (soybean, mango, banana, sugarcane, corn). Agreement ranges from 70\% to 90\%, with disagreement at or below 16\%.}
\label{fig:field_verdicts}
\end{figure}
\clearpage
\section{Agent Prompt}
\label{sec:appendix_prompt}

The agent is driven by the system prompt below, shown for the canonical full-feature configuration (reference budget $k \geq 8$, KB present, part index present). Placeholders in angle brackets are filled at runtime; lower-budget and KB-free runs adjust steps 2--5 accordingly.

\begin{verbatim}
You are an expert plant pathologist classifying a diseased plant image.

## Strategy
1. Read the test image first. Note the affected plant part
   (leaf, stem, pod, root, whole plant) and key visual features
   (color, shape, pattern, texture).
2. Read the part index file `<PART_INDEX_PATH>` and find the plant
   part you identified. This narrows the candidate classes to only
   those that affect that part. Focus on these candidates. Stay
   within the part-narrowed set. Only view classes outside it if
   you have exhausted all candidates within it and still have budget.
3. Review the symptom descriptions below to narrow further.
4. View reference images one at a time. Read ONE image, analyze how
   it compares to the test image, then decide which class to check
   next. Do NOT read multiple images in parallel. Explore before
   confirming: view one reference from EACH of your top candidates
   before viewing a second from any class.
5. IMPORTANT: Make your final prediction based on VISUAL SIMILARITY
   to reference images, not KB descriptions. The symptom descriptions
   help you understand what to look for, but when deciding between
   candidates, the reference image that most closely matches the test
   image wins. Do NOT let a text description override what you see in
   the images.
- Submit your prediction.

End your response with exactly this JSON block:
```json
{"prediction": "<class_name>", "confidence": <0.0-1.0>,
 "reasoning": "<brief explanation>"}
```
The prediction MUST be one of the provided class names (exact match).
\end{verbatim}

The accompanying user message provides the test image path, the full list of candidate class names, the per-class reference image paths with the explicit budget instruction, and the symptom KB in markdown:

\begin{verbatim}
## Test Image

Read this file to see the image you need to classify:
`<TEST_IMAGE_PATH>`

## Possible Classes (<N> total)

- <class_1>
- <class_2>
- ...

## Reference Images (Budget: exactly **<k>** reference image views.
You MUST use all <k> views before submitting. Spread across at least
**<min_classes>** different classes (1 per class first, then revisit
if budget remains).)

<total_refs> reference images. Use the Read tool to view them:
- **<class_1>**: `<path_1>`, `<path_2>`
- **<class_2>**: `<path>`
- ...

## Symptom Descriptions (Knowledge Base)

<KB markdown: per-disease symptom summary, diagnostic features,
look-alikes, with verbatim source quotes>

Now begin: Read the test image, reason through the evidence,
and end with your prediction JSON.
\end{verbatim}

The exact prompt-construction code is released with the codebase at \url{https://sage-dataset.github.io/} (\texttt{open\_agentic/prompt.py}).

\clearpage
\section{Per-Crop Accuracy Curves}
\label{sec:appendix_accuracy_curves}

Per-crop accuracy as the reference budget $k$ varies, with and without the internet KB (Sonnet). The split-radar in Section~\ref{sec:experiments} compresses the same data into a single panel; this view makes the per-$k$ trajectory explicit.

\begin{figure}[h]
\centering
\includegraphics[width=\textwidth]{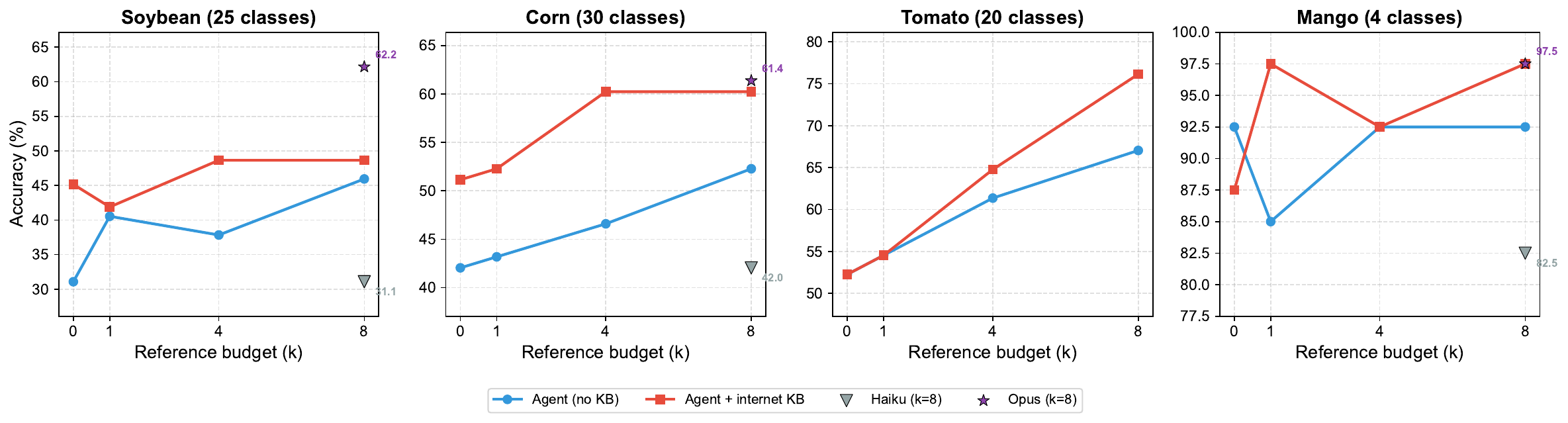}
\caption{Diagnostic accuracy vs.\ reference budget $k$ across four crops. Each panel shows the agent with no KB (blue) and with internet KB (red).}
\label{fig:accuracy_vs_k}
\end{figure}

\clearpage
\section{Few-Shot Comparison}
\label{sec:appendix_fewshot}

Table~\ref{tab:fewshot} compares the agentic pipeline against a single-pass few-shot baseline, as discussed in Section~\ref{sec:discussion}. The table covers three crops (Soybean, Corn, Mango); few-shot was not run on Tomato.

\input{tables/table_fewshot_comparison.tex}

Few-shot reaches at most $+8.1$~pp at $k{=}8$ averaged across these three crops, well below the agentic pipeline's gain at the same budget on the same crops.

\clearpage
\section{Extended Reasoning Traces}
\label{sec:appendix_traces}

The following traces show the agent's step-by-step reasoning across different crops, KB sources, models, and reference budgets. Each trace includes the configuration, the agent's observations and comparisons, and the final prediction with outcome. The traces collectively illustrate KB-guided narrowing on success cases as well as failure modes where visual ambiguity overrode KB evidence.

\input{traces/traces_appendix.tex}
\input{algo}
\end{document}

%% file: table_datasets_v2.tex
\newcommand{\yes}{\textcolor{green!55!black}{\ding{51}}}
\newcommand{\no}{\textcolor{red!60!black}{\ding{55}}}

\begin{table}[t]
\centering
\caption{Comparison of plant disease datasets and benchmarks.
\textbf{Diseases} counts disease-positive classes only (healthy/pest/weed classes excluded).
\textbf{Task}: CLS classification, Det detection, VQA visual QA, OEQ open-ended QA, MCQ multiple choice.
\textbf{Symptoms knowledge base}: \yes\ if per-disease symptom descriptions are paired with images; parenthetical notes the source.
\textbf{Demonstrated agentic eval}: \yes\ if the dataset is shown to support step-by-step diagnostic reasoning with reference image comparison.}
\label{tab:dataset_comparison}
\setlength{\tabcolsep}{4pt}
\footnotesize
\resizebox{\textwidth}{!}{%
\begin{tabular}{l r r r l c l c}
\toprule
\textbf{Dataset} & \textbf{Host species} & \textbf{Diseases} & \textbf{Images}
& \textbf{Task} & \textbf{Multi-organ} & \textbf{Symptoms knowledge base} & \textbf{Demonstrated agentic eval} \\
\midrule
PlantVillage~\cite{hughes2015open}        & 14   & 26   & 54{,}306    & CLS      & \no  & \no                          & \no \\
PlantDoc~\cite{singh2020plantdoc}         & 13   & 17   & 2{,}598     & CLS, Det & \no  & \no                          & \no \\
PlantWild~\cite{wei2024benchmarking}      & 33   & 56   & 18{,}542    & CLS      & \no  & \yes\ (LLM-generated)        & \no \\
LeafNet~\cite{quoc2026leafnet}            & 22   & 97   & 186{,}000   & VQA      & \no  & \yes\ (limited metadata)     & \no \\
CDDM~\cite{liu2024multimodal}             & 16   & 60   & 137{,}000   & MCQ      & \no  & \no                          & \no \\
AgroBench~\cite{shinoda2025agrobench}     & 203  & 682  & 3{,}745     & MCQ      & \yes & \no                          & \no \\
\midrule
\rowcolor{green!8}
\textbf{SAGE (ours)}                      & \textbf{335} & \textbf{1{,}251} & \textbf{838{,}936}
& \textbf{CLS} & \yes & \textbf{\yes\ (source-grounded)} & \yes \\
\bottomrule
\end{tabular}%
}
\end{table}

%% file: traces/trace_intext.tex
\begin{tcolorbox}[
        breakable, colback=gray!3, colframe=gray!50, boxrule=0.5pt,
        title={\footnotesize \textbf{Reasoning Trace:} Soybean Diseases — Frogeye\_leaf\_spot\_\_Soybean\_Dise\_21},
        fonttitle=\footnotesize,
    ]
    \footnotesize
    \begin{minipage}[t]{0.65\linewidth}
    \begin{tabular}{@{}ll@{}}
        \textbf{Model:} & sonnet \\
        \textbf{KB source:} & internet \\
        \textbf{Reference budget (k):} & 8 \\
        \textbf{Prediction:} & Frogeye\_leaf\_spot \\
        \textbf{Ground truth:} & Frogeye\_leaf\_spot \\
        \textbf{Outcome:} & \textcolor{green!60!black}{\textbf{Correct}} \\
        \textbf{Confidence:} & 0.93 \\
    \end{tabular}
    \end{minipage}%
    \hfill
    \begin{minipage}[t]{0.30\linewidth}
    \centering
    \fbox{\includegraphics[height=2.5cm]{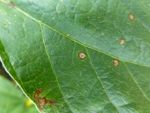}}\\[2pt]
    {\scriptsize \textit{Test image}}
    \end{minipage}

    \vspace{4pt}
    \hrule
    \vspace{4pt}

    \begin{description}[style=nextline, leftmargin=1.5cm, font=\footnotesize\bfseries]
    \item[Step 1:] \texttt{Observe test image}
    \item[Step 2:] The test image shows a soybean leaf with small, circular lesions having gray/tan centers and distinct reddish-brown margins. This pattern is very characteristic of Frogeye leaf spot. Let me check the part index and confirm with reference images.
    \item[Step 3:] \texttt{View reference: part\_index}
    \item[Step 4:] Leaf is the affected part. Top candidates: Frogeye\_leaf\_spot (gray centers, reddish-brown borders), Septoria\_brown\_spot, Phyllosticta\_leaf\_spot. Let me check references.
    \item[Step 5:] \texttt{View reference: Soybean\_Dise\_6}
    \item[Step 6:] Strong match - circular lesions with reddish-brown borders. Now checking 7 more different classes.
    \item[Step 7:] \texttt{View reference: Soybean\_Dise\_12}
    \item[Step 8:] \texttt{View reference: Soybean\_Dise\_18}
    \item[Step 9:] \texttt{View reference: Soybean\_Dise\_17}
    \item[Step 10:] \texttt{View reference: Soybean\_Dise\_74}
    \item[Step 11:] \texttt{View reference: Soybean\_Dise\_93}
    \item[Step 12:] \texttt{View reference: Soybean\_Dise\_137}
    \item[Step 13:] \texttt{View reference: Soybean\_Dise\_62}
    \item[Step 14:] \texttt{View reference: Soybean\_Dise\_39}
    \item[Step 15:] All 8 views used. The analysis is clear:

- Test image: Small circular lesions (~5-10mm), gray/tan centers, distinct reddish-brown margins, scattered on green leaf
- Frogeye\_leaf\_spot reference: Near-identical match - same circular morphology, same gray centers with reddish-brown borders
- All other candidates: Don't match - Septoria shows irregular dark lesions with yellowing; Rust shows tiny numerous specks; Cercospora shows diffuse bronze discoloration; Bacterial Blight shows angular yellow-haloed lesions; SDS shows interveinal chlorosis; Downy mildew shows pale yellow patches; Bacterial Pustule shows confluent dark blotches
    \end{description}
    \end{tcolorbox}

%% file: tables/table_main_results.tex
\begin{table*}[t]
\centering
\small
\caption{Diagnostic accuracy across crops, methods, and reference budgets $k$ (Sonnet model). Parentheses in the Crop column denote number of disease classes. The baseline is Agent (no KB) at $k{=}0$. Values show accuracy\,\% with improvement over the per-crop baseline in parentheses. Best per crop-$k$ in \textbf{bold}. Bottom rows show mean improvement (pp) over baseline, averaged across crops.}
\label{tab:main_results}
\resizebox{\textwidth}{!}{%
\begin{tabular}{llrrrr}
\toprule
Crop & Method & $k=0$ & $k=1$ & $k=4$ & $k=8$ \\
\midrule
Soybean (25) & Agent (no KB) & 31.1\% (+0.0) & 40.5\% (+9.5) & 37.8\% (+6.8) & 45.9\% (+14.9) \\
  & Agent + internet KB & \textbf{45.2\% (+14.1)} & \textbf{41.9\% (+10.8)} & \textbf{48.6\% (+17.6)} & \textbf{48.6\% (+17.6)} \\
\midrule
Corn (30) & Agent (no KB) & 42.0\% (+0.0) & 43.2\% (+1.1) & 46.6\% (+4.5) & 52.3\% (+10.2) \\
  & Agent + internet KB & \textbf{51.1\% (+9.1)} & \textbf{52.3\% (+10.2)} & \textbf{60.2\% (+18.2)} & \textbf{60.2\% (+18.2)} \\
\midrule
Tomato (20) & Agent (no KB) & \textbf{52.3\% (+0.0)} & \textbf{54.5\% (+2.3)} & 61.4\% (+9.1) & 67.0\% (+14.8) \\
  & Agent + internet KB & \textbf{52.3\% (+0.0)} & \textbf{54.5\% (+2.3)} & \textbf{64.8\% (+12.5)} & \textbf{76.1\% (+23.9)} \\
\midrule
Mango (4) & Agent (no KB) & \textbf{92.5\% (+0.0)} & 85.0\% (-7.5) & \textbf{92.5\% (+0.0)} & 92.5\% (+0.0) \\
  & Agent + internet KB & 87.5\% (-5.0) & \textbf{97.5\% (+5.0)} & \textbf{92.5\% (+0.0)} & \textbf{97.5\% (+5.0)} \\
\midrule
\textit{Mean $\Delta$ (pp)} & \textit{Agent (no KB)} & +0.0 & +1.3 & +5.1 & +10.0 \\
 & \textit{Agent + internet KB} & +4.5 & +7.1 & +12.1 & +16.2 \\
\bottomrule
\end{tabular}
}
\end{table*}

%% file: dataset_licenses_snippet.tex
\begin{footnotesize}
\begin{longtable}{@{}p{0.22\textwidth} p{0.18\textwidth} p{0.55\textwidth}@{}}
\caption{Dataset license summary: hosting and distribution rights. Each row lists the license and citation requirements for datasets included in our collection.}%
\label{tab:dataset_licenses}\\
\toprule
\textbf{Dataset} & \textbf{License} & \textbf{Notes \& Citation} \\
\midrule
\endfirsthead
\multicolumn{3}{c}{\footnotesize\itshape \tablename\ \thetable{} --- continued}\\
\toprule
\textbf{Dataset} & \textbf{License} & \textbf{Notes \& Citation} \\
\midrule
\endhead
\midrule\multicolumn{3}{r}{\footnotesize\itshape Continued on next page}\\

SBRD --- Rice Leaf Diseases &
World Bank ToU &
World Bank Dataset Terms of Use apply; review before redistribution.
Cite: Ethy, P.K., Barpanda et al.\ (2020); Ritharson, Kaggle, 2021. \\[3pt]

MangoLeafBD --- Mango Leaf Disease &
CC BY-NC 3.0 &
Attribution required; non-commercial use only.
Ahmed et al., \textit{Data in Brief} 47, 2023. DOI:~10.1016/j.dib.2023.108941. \\[3pt]

Bean Leaf Lesions Classification &
Apache 2.0 &
Attribution required; include original license text.
Marquis03, Kaggle. \\[3pt]

YellowRust19 --- Yellow Rust in Wheat &
Default license (Kaggle ToS) &
Default Kaggle license applies; contact author before redistribution.
Cite: T.\,Hayit, \textit{YellowRust19}, Kaggle, 2020. \\[3pt]

Banana Leaf Disease Dataset v1.1 &
No explicit license &
Compiled from three Kaggle datasets (rayhanarlistya, shifatearman, sujaykapadnis);
cite all original creators. \\[3pt]

Cauliflower Disease Multi Transformation &
MIT &
Attribution required; include original MIT license text.
Shuvo Kumar Basak, Kaggle. \\[3pt]

Lettuce Diseases Dataset &
Apache 2.0 &
Attribution required; include LICENSE and NOTICE file.
Ashish Jena, Kaggle. \\[3pt]

Cucumber Plant Diseases Dataset &
Default license (Kaggle ToS) &
Default Kaggle license applies; contact author before redistribution.
Cite: Karim Negm (\texttt{kareem3egm}), \textit{Cucumber Plant Diseases Dataset}, Kaggle, 2020. \\[3pt]

Durian Leaf Disease Dataset &
CC BY 4.0 &
Attribution required. Cthng123, Kaggle. \\[3pt]

Eggplant Disease Recognition Dataset &
CC BY 4.0 &
Canonical Mendeley source (DOI:~10.17632/r3tb5mzn4d.2) confirmed CC BY 4.0. Mafi \& Ava, 2023. \\[3pt]

Cotton Disease Multi Transformation &
MIT &
Attribution required; include original MIT license text.
Shuvo Kumar Basak, Kaggle. \\[3pt]

Pumpkin Leaf Disease Multi Transformation &
MIT &
Attribution required; include original MIT license text.
Shuvo Kumar Basak, Kaggle. \\[3pt]

Rose Leaf Disease Multi Transformation &
MIT &
Attribution required; include original MIT license text.
Shuvo Kumar Basak, Kaggle. \\[3pt]

Coconut Disease Multi Transformation STTV &
MIT &
Attribution required; include original MIT license text.
Shuvo Kumar Basak, Kaggle. \\[3pt]

Vanilla Plant Disease Image Dataset &
Default license (Kaggle ToS) &
Default Kaggle license applies; contact author before redistribution.
Cite: Muhammad Ihsan Permana (\texttt{mihsanpermana}), \textit{Vanilla Plant Disease Image Dataset}, Kaggle, 2024. \\[3pt]

Strawberry Disease Detection Dataset &
Default license (permissive) &
Default Kaggle license; description grants permission to use, modify, and augment.
Cite: Afzaal et al., \textit{Sensors} 21, 6565 (2021). \\[3pt]

SugarLeafIDN --- Sugarcane Leaf Diseases &
Apache 2.0 &
Attribution required; include LICENSE and NOTICE file.
Betty D.\,Puspasari, Kaggle. \\[3pt]

Cucumber Disease and Freshness (Zenodo) &
CC BY 4.0 &
Attribution required.
Cite: \textit{Cucumber Disease and Freshness Classification Dataset --- Curated Annotations},
Zenodo, 2025. DOI:~10.5281/zenodo.16816441. \\[3pt]

New Plant Diseases Dataset (Augmented) &
Default license (Kaggle ToS) &
Default Kaggle license applies; augmented derivative of PlantVillage --- contact author before redistribution
and also cite Hughes \& Salath\'{e} (PlantVillage source).
Cite: Vipoooool, \textit{New Plant Diseases Dataset}, Kaggle, 2018. \\[3pt]

FUSARIUM22 --- Fusarium Wilt in Chickpea &
Default license (Kaggle ToS) &
Default Kaggle license applies; contact author before redistribution.
Cite: Hayit et al., \textit{Eur.\ J.\ Plant Pathol.}\ 168 (2024). \\[3pt]

PlantDoc Dataset &
CC BY 4.0 &
Attribution required.
Singh et al., \textit{ACM IKDD CoDS \& COMAD}, 2020. DOI:~10.1145/3371158.3371196. \\[3pt]

Rady Plant Diseases Image-Text Pairs &
Default license (HuggingFace) &
No explicit license declared on the HuggingFace dataset card; contact author before redistribution.
Cite: Rady10, \textit{Plant-Diseases-Image-Text-Pairs}, HuggingFace Datasets. \\[3pt]

A2H0H0R1 Plant Disease (New) &
Default license (HuggingFace) &
No explicit license declared on the HuggingFace dataset card; contact author before redistribution.
Cite: A2H0H0R1, \textit{plant-disease-new}, HuggingFace Datasets. \\[3pt]

Avinash Plant Disease Classification &
Default license (HuggingFace) &
No explicit license declared on the HuggingFace dataset card; contact author before redistribution.
Cite: Avinashhm, \textit{plant-disease-classification-complete}, HuggingFace Datasets. \\[3pt]

Saketh Plant Disease &
Default license (HuggingFace) &
No explicit license declared on the HuggingFace dataset card; contact author before redistribution.
Cite: Sakethdevx, \textit{plant-disease-dataset}, HuggingFace Datasets. \\[3pt]

VQA Plant-Disease (Merged) &
Default license (HuggingFace) &
No explicit license declared on the HuggingFace dataset card; contact author before redistribution.
Cite: Raghavendrad60, \textit{vqa\_plant-disease-classification-merged-dataset}, HuggingFace Datasets. \\[3pt]

BD Crop \& Vegetable Plant Disease &
Default license (HuggingFace) &
No explicit license declared on the HuggingFace dataset card; contact author before redistribution.
Cite: Saon110, \textit{bd-crop-vegetable-plant-disease-dataset}, HuggingFace Datasets. \\[3pt]

PlantVillage Dataset &
CC BY-SA 3.0 &
Attribution + ShareAlike required; derivatives must carry the same license.
Hughes \& Salath\'{e}, arXiv:1511.08060, 2015. \\[3pt]

LeafNet Dataset &
CC BY 4.0 &
Confirmed CC BY 4.0 on HuggingFace (\texttt{enalis/LeafNet});
$\sim$70\% of data is public (training split only).
Quoc et al., arXiv:2602.13662, 2026. \\[3pt]

Soybean Stress Identification (PNAS) &
CC BY 4.0 &
Attribution required.
Ghosal et al., \textit{PNAS} 115(18), 4613--4618, 2018. DOI:~10.1073/pnas.1716999115. \\[3pt]

PlantWild v2 Dataset &
CC BY-NC-ND 4.0 &
No commercial use; no derivatives permitted.
Wei et al., \textit{ACM MM}, 2024. DOI:~10.1145/3664647.3680599. \\[3pt]

CDDM --- Crop Disease Domain Multimodal &
CC BY-NC-ND 4.0 &
No commercial use; no derivatives permitted.
Liu et al., arXiv:2503.06973, 2025. \\[3pt]

Bugwood Image Database &
Per-image attribution (Bugwood ToU) &
Image rights remain with the individual photographers/contributors;
attribution and Bugwood acknowledgement required for each image used.
Cite: The Bugwood Network, \textit{Bugwood Image Database System}, Center for Invasive Species
and Ecosystem Health, University of Georgia. \\

\end{longtable}
\end{footnotesize}

%% file: tables/table_fewshot_comparison.tex
\begin{table*}[t]
\centering
\small
\caption{Single-pass few-shot classification accuracy. The model receives $k$ randomly sampled labeled images in a single API call with no reasoning loop (Sonnet, 3 test images per class). Baseline is Agent (no KB) at $k{=}0$. Best per crop--$k$ in \textbf{bold}.}
\label{tab:fewshot}
\begin{tabular}{lrrrr}
\toprule
Crop & $k=0$ & $k=1$ & $k=4$ & $k=8$ \\
\midrule
Soybean & 41.9\% (+10.8) & 41.9\% (+10.8) & 40.5\% (+9.5) & \textbf{47.3\% (+16.2)} \\
Corn & 42.0\% (+0.0) & 38.6\% (-3.4) & 48.9\% (+6.8) & \textbf{50.0\% (+8.0)} \\
Mango & 92.5\% (+0.0) & \textbf{95.0\% (+2.5)} & 92.5\% (+0.0) & 92.5\% (+0.0) \\
\midrule
\textit{Mean $\Delta$ (pp)} & \textit{+3.6} & \textit{+3.3} & \textit{+5.4} & \textit{+8.1} \\
\bottomrule
\end{tabular}
\end{table*}

%% file: traces/traces_appendix.tex

\input{traces/trace_appendix_01.tex}
\vspace{8pt}
\input{traces/trace_appendix_02.tex}
\vspace{8pt}
\input{traces/trace_appendix_03.tex}
\vspace{8pt}
\input{traces/trace_appendix_04.tex}
\vspace{8pt}
\input{traces/trace_appendix_05.tex}
\vspace{8pt}
\input{traces/trace_appendix_06.tex}
\vspace{8pt}
\input{traces/trace_appendix_07.tex}
\vspace{8pt}
\input{traces/trace_appendix_08.tex}
\vspace{8pt}
\input{traces/trace_appendix_09.tex}
\vspace{8pt}
\input{traces/trace_appendix_10.tex}
\vspace{8pt}
\input{traces/trace_appendix_11.tex}
\vspace{8pt}
\input{traces/trace_appendix_12.tex}
\vspace{8pt}
\input{traces/trace_appendix_13.tex}
\vspace{8pt}
\input{traces/trace_appendix_14.tex}
\vspace{8pt}
\input{traces/trace_appendix_15.tex}
\vspace{8pt}
\input{traces/trace_appendix_16.tex}

%% file: traces/trace_appendix_01.tex
\begin{tcolorbox}[
        breakable, colback=gray!3, colframe=gray!50, boxrule=0.5pt,
        title={\footnotesize \textbf{Reasoning Trace:} Soybean Diseases — Bacterial\_Blight\_\_Soybean\_Dise\_125},
        fonttitle=\footnotesize,
    ]
    \footnotesize
    \begin{minipage}[t]{0.65\linewidth}
    \begin{tabular}{@{}ll@{}}
        \textbf{Model:} & sonnet \\
        \textbf{KB source:} & internet \\
        \textbf{Reference budget (k):} & 4 \\
        \textbf{Prediction:} & Bacterial\_Blight \\
        \textbf{Ground truth:} & Bacterial\_Blight \\
        \textbf{Outcome:} & \textcolor{green!60!black}{\textbf{Correct}} \\
        \textbf{Confidence:} & 0.82 \\
    \end{tabular}
    \end{minipage}%
    \hfill
    \begin{minipage}[t]{0.30\linewidth}
    \centering
    \fbox{\includegraphics[height=2.5cm]{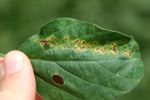}}\\[2pt]
    {\scriptsize \textit{Test image}}
    \end{minipage}

    \vspace{4pt}
    \hrule
    \vspace{4pt}

    \begin{description}[style=nextline, leftmargin=1.5cm, font=\footnotesize\bfseries]
    \item[Step 1:] \texttt{Observe test image}
    \item[Step 2:] The test image shows a soybean leaf with dark brown/black irregular lesions surrounded by yellow/chlorotic halos. The lesions appear angular and concentrated along the mid-vein area. The green tissue between lesions remains intact.

Top candidates: Bacterial\_Blight (angular lesions + yellow halos), Septoria\_brown\_spot (dark irregular lesions), Soybean\_Vein\_necrosis\_virus (lesions along veins with yellow tissue).

Now I'll view one reference from each top candidate.
    \item[Step 3:] \texttt{View reference: Soybean\_Dise\_17}
    \item[Step 4:] This Bacterial\_Blight reference matches well - dark irregular lesions with yellow-green halos. Let me check Septoria\_brown\_spot next.
    \item[Step 5:] \texttt{View reference: Soybean\_Dise\_12}
    \item[Step 6:] Septoria shows extensive yellowing with small dark spots - different from the test image which has a mostly green leaf with distinct dark lesions and yellow halos. Let me check Soybean\_Vein\_necrosis\_virus.
    \item[Step 7:] \texttt{View reference: Soybean\_Dise\_126}
    \item[Step 8:] Vein necrosis shows diffuse yellow spots without clear dark lesion centers - doesn't match. I need one more class view. The dark lesions with yellow halos in the test image best match Bacterial\_Blight so far. Let me check Frogeye\_leaf\_spot as my 4th class.
    \item[Step 9:] \texttt{View reference: Soybean\_Dise\_6}
    \item[Step 10:] Analysis summary:

- Test image: Green soybean leaf with dark brown/black irregular lesions surrounded by yellow-green halos, concentrated along the mid-vein.
- Bacterial\_Blight reference: Nearly identical - dark angular lesions with yellow-green halos on green leaf. Best match.
- Septoria\_brown\_spot: Extensive yellowing with tiny dark spots - doesn't match (test leaf is mostly green).
- Soybean\_Vein\_necrosis\_virus: Diffuse yellowish spots without distinct dark centers - doesn't match.
- Frogeye\_leaf\_spot: Small circular lesions with holes, reddish-brown borders - doesn't match the larger irregular pattern.

The yellow-green halos + dark angular/irregular lesions on an otherwise green leaf is the hallmark of Bacterial\_Blight.
    \end{description}
    \end{tcolorbox}

%% file: traces/trace_appendix_02.tex
\begin{tcolorbox}[
        breakable, colback=gray!3, colframe=gray!50, boxrule=0.5pt,
        title={\footnotesize \textbf{Reasoning Trace:} Soybean Diseases — Anthracnose\_\_Soybean\_Dise\_10},
        fonttitle=\footnotesize,
    ]
    \footnotesize
    \begin{minipage}[t]{0.65\linewidth}
    \begin{tabular}{@{}ll@{}}
        \textbf{Model:} & sonnet \\
        \textbf{KB source:} & internet \\
        \textbf{Reference budget (k):} & 4 \\
        \textbf{Prediction:} & Charcoal\_Rot \\
        \textbf{Ground truth:} & Anthracnose \\
        \textbf{Outcome:} & \textcolor{red}{\textbf{Incorrect}} \\
        \textbf{Confidence:} & 0.88 \\
    \end{tabular}
    \end{minipage}%
    \hfill
    \begin{minipage}[t]{0.30\linewidth}
    \centering
    \fbox{\includegraphics[height=2.5cm]{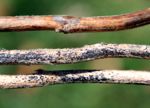}}\\[2pt]
    {\scriptsize \textit{Test image}}
    \end{minipage}

    \vspace{4pt}
    \hrule
    \vspace{4pt}

    \begin{description}[style=nextline, leftmargin=1.5cm, font=\footnotesize\bfseries]
    \item[Step 1:] \texttt{Observe test image}
    \item[Step 2:] Key observations from the test image:
- Plant part: Stem (3 stems shown close-up)
- Key features: Bleached/pale white-tan stems with numerous small black specks scattered across the surface
- Texture: Fluffy/hairy white mycelial growth visible on stems
- Black structures: Many small black bodies embedded in/on the white bleached stem tissue

The white bleached stems + black sclerotia embedded in white mycelium is classic White Mold. Let me verify with reference images, then also check Charcoal Rot (which has black microsclerotia under the epidermis) as a candidate.
    \item[Step 3:] \texttt{View reference: Soybean\_Dise\_339}
    \item[Step 4:] This White Mold reference shows early-stage large sclerotia but the stem is still green. Let me check Charcoal Rot to compare the black speck pattern.
    \item[Step 5:] \texttt{View reference: Soybean\_Dise\_108}
    \item[Step 6:] Charcoal Rot shows a pale stem but without the dense black specks pattern. Now let me check the other White Mold stem image.
    \item[Step 7:] \texttt{View reference: Soybean\_Dise\_56}
    \item[Step 8:] Both White Mold references show bleached stems with black structures, but the test image has very dense, numerous tiny black specks — more consistent with microsclerotia. Let me use my last view on Charcoal Rot to compare.
    \item[Step 9:] \texttt{View reference: Soybean\_Dise\_49}
    \item[Step 10:] This is a strong match. The Charcoal Rot reference (Soybean\_Dise\_49) shows exactly the same pattern as the test image:
- Multiple stems shown together
- Pale/bleached tan-gray stems
- Dense, numerous tiny black specks scattered across the stem surface (microsclerotia giving the "charcoal-sprinkled" appearance)
- The internal split stem shows white pith

This matches the test image almost perfectly — three stems displayed, pale/bleached with countless tiny black dots (microsclerotia) across the surface. White Mold sclerotia are larger and fewer; what's visible here is the characteristic microsclerotia of Charcoal Rot.
    \end{description}
    \end{tcolorbox}

%% file: traces/trace_appendix_03.tex
\begin{tcolorbox}[
        breakable, colback=gray!3, colframe=gray!50, boxrule=0.5pt,
        title={\footnotesize \textbf{Reasoning Trace:} Soybean Diseases — Bacterial\_Blight\_\_Soybean\_Dise\_125},
        fonttitle=\footnotesize,
    ]
    \footnotesize
    \begin{minipage}[t]{0.65\linewidth}
    \begin{tabular}{@{}ll@{}}
        \textbf{Model:} & sonnet \\
        \textbf{KB source:} & no KB \\
        \textbf{Reference budget (k):} & 4 \\
        \textbf{Prediction:} & Bacterial\_Blight \\
        \textbf{Ground truth:} & Bacterial\_Blight \\
        \textbf{Outcome:} & \textcolor{green!60!black}{\textbf{Correct}} \\
        \textbf{Confidence:} & 0.87 \\
    \end{tabular}
    \end{minipage}%
    \hfill
    \begin{minipage}[t]{0.30\linewidth}
    \centering
    \fbox{\includegraphics[height=2.5cm]{traces/images/Bacterial_Blight__Soybean_Dise_125.jpg}}\\[2pt]
    {\scriptsize \textit{Test image}}
    \end{minipage}

    \vspace{4pt}
    \hrule
    \vspace{4pt}

    \begin{description}[style=nextline, leftmargin=1.5cm, font=\footnotesize\bfseries]
    \item[Step 1:] \texttt{Observe test image}
    \item[Step 2:] The test image shows a soybean leaf with dark brown/purplish-black lesions surrounded by yellow halos, distributed along the leaf midrib area. There's also one hole. The yellow halo + dark center pattern strongly suggests bacterial or fungal leaf spotting.

Top candidates: Bacterial\_Blight (angular dark lesions + yellow halos), Septoria\_brown\_spot, Cercospora, Frogeye\_leaf\_spot.

Now viewing one reference from each of my 4 top candidates:
    \item[Step 3:] \texttt{View reference: Soybean\_Dise\_17}
    \item[Step 4:] This Bacterial\_Blight reference matches very well - dark brown/purple lesions with yellow halos. Now checking Septoria and Cercospora.
    \item[Step 5:] \texttt{View reference: Soybean\_Dise\_12}
    \item[Step 6:] \texttt{View reference: Soybean\_Dise\_33}
    \item[Step 7:] The Cercospora reference shows seed discoloration - completely different from the test image. The Bacterial\_Blight reference is a clear match.

Analysis:
- Test image: soybean leaf with dark brown/purple lesions + prominent yellow halos, angular distribution along midrib
- Bacterial\_Blight reference: identical pattern - dark lesions with yellow halos on green leaf tissue
- Septoria\_brown\_spot: more extensive yellowing, smaller spots scattered throughout
- Cercospora: seed discoloration, not leaf lesions

The yellow halo + dark angular lesion pattern is the defining feature of Bacterial\_Blight.
    \end{description}
    \end{tcolorbox}

%% file: traces/trace_appendix_04.tex
\begin{tcolorbox}[
        breakable, colback=gray!3, colframe=gray!50, boxrule=0.5pt,
        title={\footnotesize \textbf{Reasoning Trace:} Soybean Diseases — Anthracnose\_\_Soybean\_Dise\_10},
        fonttitle=\footnotesize,
    ]
    \footnotesize
    \begin{minipage}[t]{0.65\linewidth}
    \begin{tabular}{@{}ll@{}}
        \textbf{Model:} & sonnet \\
        \textbf{KB source:} & no KB \\
        \textbf{Reference budget (k):} & 4 \\
        \textbf{Prediction:} & Phomopsis \\
        \textbf{Ground truth:} & Anthracnose \\
        \textbf{Outcome:} & \textcolor{red}{\textbf{Incorrect}} \\
        \textbf{Confidence:} & 0.62 \\
    \end{tabular}
    \end{minipage}%
    \hfill
    \begin{minipage}[t]{0.30\linewidth}
    \centering
    \fbox{\includegraphics[height=2.5cm]{traces/images/Anthracnose__Soybean_Dise_10.jpg}}\\[2pt]
    {\scriptsize \textit{Test image}}
    \end{minipage}

    \vspace{4pt}
    \hrule
    \vspace{4pt}

    \begin{description}[style=nextline, leftmargin=1.5cm, font=\footnotesize\bfseries]
    \item[Step 1:] \texttt{Observe test image}
    \item[Step 2:] The test image shows soybean stems with pale/bleached (cream-white) coloration and scattered small black dots across the stem surface. These are likely pycnidia or microsclerotia. Candidates: Phomopsis (bleached stems + black pycnidia), Charcoal\_Rot (gray stem + black microsclerotia), Anthracnose (dark lesions + black acervuli).

Now viewing one reference from each top candidate:
    \item[Step 3:] \texttt{View reference: Soybean\_Dise\_14}
    \item[Step 4:] This Phomopsis image shows a pod. Let me check Charcoal\_Rot and Anthracnose references.
    \item[Step 5:] \texttt{View reference: Soybean\_Dise\_10}
    \item[Step 6:] Let me use my last view on another Charcoal\_Rot reference showing the stem close-up:
    \item[Step 7:] \texttt{View reference: Soybean\_Dise\_52}
    \item[Step 8:] Visual analysis of the test image:
- 3 soybean stems shown close-up
- Stems are pale/cream/bleached white
- Numerous small black dots scattered across the stem surface
- Top stem is more brownish (earlier stage?), middle and bottom stems are distinctly bleached

Reasoning:
The characteristic pale/cream bleached stems with scattered black pycnidia (fruiting bodies) visible on the surface is classic Phomopsis stem blight. The bleaching pattern and black dot distribution match this disease. Charcoal\_Rot was also considered (microsclerotia + silvery-gray stem), but Charcoal\_Rot typically shows more grayish discoloration rather than this cream/white bleaching. The reference Charcoal\_Rot images also showed field-level wilting symptoms rather than close-up stem bleaching with black dots.
    \end{description}
    \end{tcolorbox}

%% file: traces/trace_appendix_05.tex
\begin{tcolorbox}[
        breakable, colback=gray!3, colframe=gray!50, boxrule=0.5pt,
        title={\footnotesize \textbf{Reasoning Trace:} Soybean Diseases — Bacterial\_Blight\_\_Soybean\_Dise\_125},
        fonttitle=\footnotesize,
    ]
    \footnotesize
    \begin{minipage}[t]{0.65\linewidth}
    \begin{tabular}{@{}ll@{}}
        \textbf{Model:} & opus \\
        \textbf{KB source:} & internet \\
        \textbf{Reference budget (k):} & 8 \\
        \textbf{Prediction:} & Bacterial\_Blight \\
        \textbf{Ground truth:} & Bacterial\_Blight \\
        \textbf{Outcome:} & \textcolor{green!60!black}{\textbf{Correct}} \\
        \textbf{Confidence:} & 0.92 \\
    \end{tabular}
    \end{minipage}%
    \hfill
    \begin{minipage}[t]{0.30\linewidth}
    \centering
    \fbox{\includegraphics[height=2.5cm]{traces/images/Bacterial_Blight__Soybean_Dise_125.jpg}}\\[2pt]
    {\scriptsize \textit{Test image}}
    \end{minipage}

    \vspace{4pt}
    \hrule
    \vspace{4pt}

    \begin{description}[style=nextline, leftmargin=1.5cm, font=\footnotesize\bfseries]
    \item[Step 1:] \texttt{Observe test image}
    \item[Step 2:] \texttt{View reference: part\_index}
    \item[Step 3:] The test image shows a soybean leaf with dark brown irregular lesions concentrated in a band, surrounded by yellowish-green halos. Some tissue has fallen out leaving holes, giving a ragged appearance. Lesions appear somewhat angular.

This is a leaf disease. Top candidates based on symptoms: Bacterial\_Blight, Septoria\_brown\_spot, Soybean\_Vein\_necrosis\_virus, Bacterial\_Pustule. Let me view references across 8 classes.
    \item[Step 4:] \texttt{View reference: Soybean\_Dise\_17}
    \item[Step 5:] Bacterial\_Blight reference shows angular brown spots with yellow halos, merged dead patches. Very similar pattern to test image. Let me check other candidates.
    \item[Step 6:] \texttt{View reference: Soybean\_Dise\_12}
    \item[Step 7:] Septoria brown spot shows numerous small dark spots with extensive leaf yellowing. Different pattern from the test image. Let me continue.
    \item[Step 8:] \texttt{View reference: Soybean\_Dise\_126}
    \item[Step 9:] SVN shows yellow spots along veins, different from the test image's angular merged lesions with halos. Let me check more.
    \item[Step 10:] \texttt{View reference: Soybean\_Dise\_11}
    \item[Step 11:] Bacterial Pustule reference shows a close-up of severely affected leaf tissue with extensive brown/orange discoloration. Very different from the test image. Let me check more classes.
    \item[Step 12:] \texttt{View reference: Soybean\_Dise\_6}
    \item[Step 13:] Frogeye leaf spot shows discrete circular gray spots with reddish-brown borders, scattered across the leaf. Very different from the test image's merged angular lesions. Let me continue.
    \item[Step 14:] \texttt{View reference: Soybean\_Dise\_137}
    \item[Step 15:] Downy mildew shows light yellow spots spread across the leaf surface. Very different from the test image. Let me check two more classes.
    \item[Step 16:] \texttt{View reference: Soybean\_Dise\_74}
    \item[Step 17:] Soybean rust shows small tan to brick-red spots scattered across the leaf. Very different from the test image. One more class to check.
    \item[Step 18:] \texttt{View reference: Soybean\_Dise\_18}
    \item[Step 19:] Phyllosticta shows large tan lesions with dark margins, different pattern from the test image.

I've now viewed 8 references across 8 different classes. The evidence strongly points to Bacterial\_Blight:

- Angular, vein-limited brown lesions matching the description perfectly
- Yellowish-green halos surrounding the darkened spots
- Merged dead patches where spots have coalesced
- Tissue falling out leaving holes, creating the characteristic "ragged appearance"
- The Bacterial\_Blight reference image showed the most similar pattern

No other class matched: Septoria had extensive yellowing with small scattered spots, Frogeye had discrete circular gray-centered spots, Downy mildew/Soybean rust had very different spot patterns, and Phyllosticta had larger irregular tan lesions.
    \end{description}
    \end{tcolorbox}

%% file: traces/trace_appendix_06.tex
\begin{tcolorbox}[
        breakable, colback=gray!3, colframe=gray!50, boxrule=0.5pt,
        title={\footnotesize \textbf{Reasoning Trace:} Soybean Diseases — Anthracnose\_\_Soybean\_Dise\_10},
        fonttitle=\footnotesize,
    ]
    \footnotesize
    \begin{minipage}[t]{0.65\linewidth}
    \begin{tabular}{@{}ll@{}}
        \textbf{Model:} & opus \\
        \textbf{KB source:} & internet \\
        \textbf{Reference budget (k):} & 8 \\
        \textbf{Prediction:} & Charcoal\_Rot \\
        \textbf{Ground truth:} & Anthracnose \\
        \textbf{Outcome:} & \textcolor{red}{\textbf{Incorrect}} \\
        \textbf{Confidence:} & 0.92 \\
    \end{tabular}
    \end{minipage}%
    \hfill
    \begin{minipage}[t]{0.30\linewidth}
    \centering
    \fbox{\includegraphics[height=2.5cm]{traces/images/Anthracnose__Soybean_Dise_10.jpg}}\\[2pt]
    {\scriptsize \textit{Test image}}
    \end{minipage}

    \vspace{4pt}
    \hrule
    \vspace{4pt}

    \begin{description}[style=nextline, leftmargin=1.5cm, font=\footnotesize\bfseries]
    \item[Step 1:] \texttt{Observe test image}
    \item[Step 2:] \texttt{View reference: part\_index}
    \item[Step 3:] The test image shows soybean stems with numerous small black specks/dots scattered across bleached/tan stem surfaces. This narrows candidates to stem diseases.

Top candidates based on symptoms:
- Charcoal\_Rot: microsclerotia giving "charcoal-sprinkled" appearance
- Anthracnose: black acervuli with setae on stems

Let me view references systematically, one at a time.
    \item[Step 4:] \texttt{View reference: Soybean\_Dise\_108}
    \item[Step 5:] Charcoal rot reference shows gray streaks but not the dense black specks pattern in the test image. Let me check Anthracnose.
    \item[Step 6:] \texttt{View reference: Soybean\_Dise\_17}
    \item[Step 7:] Anthracnose shows dark lesions on stems with pods attached. Let me check the other Charcoal Rot stem reference.
    \item[Step 8:] \texttt{View reference: Soybean\_Dise\_49}
    \item[Step 9:] This Charcoal Rot reference is a strong match - small black specks (microsclerotia) scattered on gray/tan stems, very similar to the test image. Let me check other stem disease candidates for comparison.
    \item[Step 10:] \texttt{View reference: Soybean\_Dise\_339}
    \item[Step 11:] White Mold shows fluffy white mycelium and large black sclerotia - doesn't match. Let me check Anthracnose stem more closely and other candidates.
    \item[Step 12:] \texttt{View reference: Soybean\_Dise\_67}
    \item[Step 13:] Anthracnose on green stems with pods - different stage and appearance. Let me check Phytophthora stem and Rhizoctonia stem.
    \item[Step 14:] \texttt{View reference: Soybean\_Dise\_57}
    \item[Step 15:] Phytophthora shows dark chocolate-brown stem discoloration on living plants - doesn't match. Let me check Rhizoctonia and Fusarium stems.
    \item[Step 16:] \texttt{View reference: Soybean\_Dise\_18}
    \item[Step 17:] Rhizoctonia shows sunken lesions at the soil line on seedlings - doesn't match.
    \item[Step 18:] \texttt{View reference: Soybean\_Dise\_54}
    \item[Step 19:] Fusarium shows discolored roots/stems on seedlings - doesn't match.

All 8 reference views used. The evidence strongly points to Charcoal Rot:

- Test image: Mature, dry soybean stems with dense, small black specks scattered across a gray/tan stem surface
- Charcoal Rot reference (Dise\_49): Nearly identical pattern of small black microsclerotia on gray stems
- Symptom description match: "small black specks (microsclerotia) beneath the epidermis giving a charcoal-sprinkled appearance" matches perfectly
- Other candidates eliminated: Anthracnose (different fruiting body pattern), White Mold (fluffy mycelium + large sclerotia), Phytophthora (chocolate-brown discoloration on living stems), Rhizoctonia (soil-line lesions), Fusarium (seedling stage)
    \end{description}
    \end{tcolorbox}

%% file: traces/trace_appendix_07.tex
\begin{tcolorbox}[
        breakable, colback=gray!3, colframe=gray!50, boxrule=0.5pt,
        title={\footnotesize \textbf{Reasoning Trace:} Corn Diseases — Anthracnose\_Leaf\_Spot\_And\_Top\_Dieback\_\_Corn\_Disease\_47},
        fonttitle=\footnotesize,
    ]
    \footnotesize
    \begin{minipage}[t]{0.65\linewidth}
    \begin{tabular}{@{}ll@{}}
        \textbf{Model:} & sonnet \\
        \textbf{KB source:} & internet \\
        \textbf{Reference budget (k):} & 4 \\
        \textbf{Prediction:} & Anthracnose\_Leaf\_Spot\_And\_Top\_Dieback \\
        \textbf{Ground truth:} & Anthracnose\_Leaf\_Spot\_And\_Top\_Dieback \\
        \textbf{Outcome:} & \textcolor{green!60!black}{\textbf{Correct}} \\
        \textbf{Confidence:} & 0.42 \\
    \end{tabular}
    \end{minipage}%
    \hfill
    \begin{minipage}[t]{0.30\linewidth}
    \centering
    \fbox{\includegraphics[height=2.5cm]{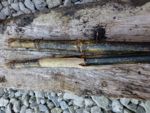}}\\[2pt]
    {\scriptsize \textit{Test image}}
    \end{minipage}

    \vspace{4pt}
    \hrule
    \vspace{4pt}

    \begin{description}[style=nextline, leftmargin=1.5cm, font=\footnotesize\bfseries]
    \item[Step 1:] \texttt{Observe test image}
    \item[Step 2:] \texttt{View reference: part\_index}
    \item[Step 3:] The test image shows corn stalks with very dark blackish/bluish-black discoloration, particularly on the rind, with yellowish areas near nodes. This is a stem specimen. Stem candidates: Anthracnose, Bacterial\_Stalk\_Rot, Charcoal\_Stalk\_Rot, Diplodia\_Stalk\_Rot, Physoderma\_Stalk\_Rot, Purple\_Leaf\_Sheath.

The dark blackish nodes are striking. Top candidates: Physoderma\_Stalk\_Rot (black nodes), Anthracnose (black shiny stalk), Charcoal\_Stalk\_Rot (charred interior), Diplodia\_Stalk\_Rot. Let me view one reference per top candidate.
    \item[Step 4:] \texttt{View reference: Corn\_Disease\_17}
    \item[Step 5:] \texttt{View reference: Corn\_Disease\_12}
    \item[Step 6:] \texttt{View reference: Corn\_Disease\_10}
    \item[Step 7:] \texttt{View reference: Corn\_Disease\_2}
    \item[Step 8:] Test image analysis:
- Affected part: stem (stalk sections cut and laid on wooden planks)
- Key features: extensive deep blue-black discoloration across the entire stalk rind, yellowish-tan coloring at nodes, multiple stalk sections

Reference comparison:
1. Physoderma\_Stalk\_Rot — reference shows localized dark spots on green stalk. Doesn't match the extensive rind blackening.
2. Charcoal\_Stalk\_Rot — reference shows dark shredded interior with charred pith. Interior charring doesn't match the exterior rind blackening in the test image.
3. Anthracnose (stem) — reference shows brownish-black patches on rind with some rotted sections. Closer match — Anthracnose stalk rot is known for extensive black discoloration of the stalk rind.
4. Purple\_Leaf\_Sheath — reference shows localized dark lesion on otherwise green stalk. Doesn't match.

The deep, extensive blue-black coloration of the entire stalk exterior rind, with yellowish nodes, is most consistent with Anthracnose stalk rot. The Anthracnose stalk phase causes characteristic shiny black lesions across the lower stalk rind — which aligns with what's visible.
    \end{description}
    \end{tcolorbox}

%% file: traces/trace_appendix_08.tex
\begin{tcolorbox}[
        breakable, colback=gray!3, colframe=gray!50, boxrule=0.5pt,
        title={\footnotesize \textbf{Reasoning Trace:} Corn Diseases — Aspergillus\_Ear\_Rot\_\_Corn\_Disease\_19},
        fonttitle=\footnotesize,
    ]
    \footnotesize
    \begin{minipage}[t]{0.65\linewidth}
    \begin{tabular}{@{}ll@{}}
        \textbf{Model:} & sonnet \\
        \textbf{KB source:} & internet \\
        \textbf{Reference budget (k):} & 4 \\
        \textbf{Prediction:} & Fusarium\_Disease \\
        \textbf{Ground truth:} & Aspergillus\_Ear\_Rot \\
        \textbf{Outcome:} & \textcolor{red}{\textbf{Incorrect}} \\
        \textbf{Confidence:} & 0.50 \\
    \end{tabular}
    \end{minipage}%
    \hfill
    \begin{minipage}[t]{0.30\linewidth}
    \centering
    \fbox{\includegraphics[height=2.5cm]{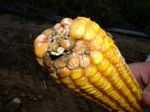}}\\[2pt]
    {\scriptsize \textit{Test image}}
    \end{minipage}

    \vspace{4pt}
    \hrule
    \vspace{4pt}

    \begin{description}[style=nextline, leftmargin=1.5cm, font=\footnotesize\bfseries]
    \item[Step 1:] \texttt{Observe test image}
    \item[Step 2:] \texttt{View reference: part\_index}
    \item[Step 3:] The test image shows a corn ear with damage concentrated at the tip — brownish/reddish-pink rotted kernels at the tip, some whitish material mixed in, healthy yellow kernels below. This is a seed/pod part. Top candidates: Fusarium\_Disease, Gibberella\_Disease, Aspergillus\_Ear\_Rot, Diplodia\_Ear\_Rot.
    \item[Step 4:] \texttt{View reference: Corn\_Disease\_25}
    \item[Step 5:] \texttt{View reference: Corn\_Disease\_18}
    \item[Step 6:] \texttt{View reference: Corn\_Disease\_15}
    \item[Step 7:] \texttt{View reference: Corn\_Disease\_27}
    \item[Step 8:] Analysis:

- Test image: Corn ear with tip rot — rotted/shriveled brownish-reddish-pink kernels concentrated at the tip, whitish material in the damaged area, insect damage visible, healthy yellow kernels below.
- Fusarium\_Disease ref: Whitish-pinkish mold scattered mid-ear — partial match (tip location fits, color fits partially).
- Gibberella\_Disease ref: Vivid red/pink mold starting at tip — the red color is more intense than the test image's brownish-pink.
- Aspergillus\_Ear\_Rot ref: Olive-green powdery mold — no match.
- Diplodia\_Ear\_Rot ref: White mycelium from base, bleached husks — no match.

Reasoning: The test image shows tip-concentrated rot with brownish-pinkish kernels and some whitish mold material. Fusarium is described as "whitish to pink-colored fungal growth on kernels, often at the ear tip" with association with insect damage entry points. The whitish material visible + tip location + pinkish-brownish coloration fits Fusarium better than the vivid red of Gibberella. Aspergillus and Diplodia are clearly ruled out.
    \end{description}
    \end{tcolorbox}

%% file: traces/trace_appendix_09.tex
\begin{tcolorbox}[
        breakable, colback=gray!3, colframe=gray!50, boxrule=0.5pt,
        title={\footnotesize \textbf{Reasoning Trace:} Corn Diseases — Anthracnose\_Leaf\_Spot\_And\_Top\_Dieback\_\_Corn\_Disease\_47},
        fonttitle=\footnotesize,
    ]
    \footnotesize
    \begin{minipage}[t]{0.65\linewidth}
    \begin{tabular}{@{}ll@{}}
        \textbf{Model:} & sonnet \\
        \textbf{KB source:} & no KB \\
        \textbf{Reference budget (k):} & 8 \\
        \textbf{Prediction:} & Anthracnose\_Leaf\_Spot\_And\_Top\_Dieback \\
        \textbf{Ground truth:} & Anthracnose\_Leaf\_Spot\_And\_Top\_Dieback \\
        \textbf{Outcome:} & \textcolor{green!60!black}{\textbf{Correct}} \\
        \textbf{Confidence:} & 0.52 \\
    \end{tabular}
    \end{minipage}%
    \hfill
    \begin{minipage}[t]{0.30\linewidth}
    \centering
    \fbox{\includegraphics[height=2.5cm]{traces/images/Anthracnose_Leaf_Spot_And_Top_Dieback__Corn_Disease_47.jpg}}\\[2pt]
    {\scriptsize \textit{Test image}}
    \end{minipage}

    \vspace{4pt}
    \hrule
    \vspace{4pt}

    \begin{description}[style=nextline, leftmargin=1.5cm, font=\footnotesize\bfseries]
    \item[Step 1:] \texttt{Observe test image}
    \item[Step 2:] The test image shows corn stalks laid out, with dark/black discoloration on the outer surface and yellow-tan internal tissue. This looks like a stalk rot disease. Key features: dark blackened exterior, yellowing/tan mixed tissue, stalks split open showing internal decay.

Candidates: Charcoal\_Stalk\_Rot, Diplodia\_Stalk\_Rot, Gibberella\_Disease, Bacterial\_Stalk\_Rot, Physoderma\_Stalk\_Rot.

Let me view reference images one at a time.
    \item[Step 3:] \texttt{View reference: Corn\_Disease\_12}
    \item[Step 4:] \texttt{View reference: Corn\_Disease\_14}
    \item[Step 5:] \texttt{View reference: Corn\_Disease\_17}
    \item[Step 6:] \texttt{View reference: Corn\_Disease\_12}
    \item[Step 7:] \texttt{View reference: Corn\_Disease\_16}
    \item[Step 8:] \texttt{View reference: Corn\_Disease\_32}
    \item[Step 9:] \texttt{View reference: Corn\_Disease\_10}
    \item[Step 10:] \texttt{View reference: Corn\_Disease\_52}
    \item[Step 11:] I've now viewed 8 reference images. Let me assess:

Test image features: Two corn stalks laid out on a wooden surface, showing heavy dark/black exterior discoloration with yellow-tan patches, one stalk split open revealing fibrous internal tissue. Classic stalk rot presentation.

Reference comparison:
- Charcoal\_Stalk\_Rot (Corn\_Disease\_12, 32): Shows dark internal pith with gray-black sclerotia — internal focus, doesn't match the heavy black exterior
- Diplodia\_Stalk\_Rot (Corn\_Disease\_14): Dry, tan/brown stalk breaking at node — no black exterior
- Physoderma\_Stalk\_Rot (Corn\_Disease\_17): Dark spots on a green stalk — very different
- Gibberella\_Disease (Corn\_Disease\_12): Ear rot with pink/white mold — completely different
- Bacterial\_Stalk\_Rot (Corn\_Disease\_16): Tan/brown leaf sheath discoloration — different
- Anthracnose\_Leaf\_Spot\_And\_Top\_Dieback (Corn\_Disease\_10): Shows stalks with brown/dark exterior patches including black areas at nodes — closer match. Anthracnose stalk rot is specifically known for causing shiny black discoloration of the outer stalk rind

The heavy black/dark exterior discoloration on the stalks is most consistent with Anthracnose stalk rot, which is part of the Anthracnose\_Leaf\_Spot\_And\_Top\_Dieback complex. The Corn\_Disease\_10 reference showed similar dark stalk exterior features.
    \end{description}
    \end{tcolorbox}

%% file: traces/trace_appendix_10.tex
\begin{tcolorbox}[
        breakable, colback=gray!3, colframe=gray!50, boxrule=0.5pt,
        title={\footnotesize \textbf{Reasoning Trace:} Corn Diseases — Aspergillus\_Ear\_Rot\_\_Corn\_Disease\_19},
        fonttitle=\footnotesize,
    ]
    \footnotesize
    \begin{minipage}[t]{0.65\linewidth}
    \begin{tabular}{@{}ll@{}}
        \textbf{Model:} & sonnet \\
        \textbf{KB source:} & no KB \\
        \textbf{Reference budget (k):} & 8 \\
        \textbf{Prediction:} & Gibberella\_Disease \\
        \textbf{Ground truth:} & Aspergillus\_Ear\_Rot \\
        \textbf{Outcome:} & \textcolor{red}{\textbf{Incorrect}} \\
        \textbf{Confidence:} & 0.82 \\
    \end{tabular}
    \end{minipage}%
    \hfill
    \begin{minipage}[t]{0.30\linewidth}
    \centering
    \fbox{\includegraphics[height=2.5cm]{traces/images/Aspergillus_Ear_Rot__Corn_Disease_19.jpg}}\\[2pt]
    {\scriptsize \textit{Test image}}
    \end{minipage}

    \vspace{4pt}
    \hrule
    \vspace{4pt}

    \begin{description}[style=nextline, leftmargin=1.5cm, font=\footnotesize\bfseries]
    \item[Step 1:] \texttt{Observe test image}
    \item[Step 2:] The test image shows a corn ear with rot concentrated at the tip. The affected kernels are brown/pinkish with some white-gray mold growth. The rest of the ear has healthy yellow kernels. This pattern suggests an ear rot disease — likely Gibberella (pink/red tip rot) or Fusarium.

Let me check reference images for top candidates.
    \item[Step 3:] \texttt{View reference: Corn\_Disease\_12}
    \item[Step 4:] \texttt{View reference: Corn\_Disease\_25}
    \item[Step 5:] \texttt{View reference: Corn\_Disease\_27}
    \item[Step 6:] \texttt{View reference: Corn\_Disease\_15}
    \item[Step 7:] \texttt{View reference: Corn\_Disease\_42}
    \item[Step 8:] \texttt{View reference: Corn\_Disease\_2}
    \item[Step 9:] \texttt{View reference: Corn\_Disease\_42}
    \item[Step 10:] \texttt{View reference: Corn\_Disease\_64}
    \item[Step 11:] Analysis summary:

- Test image: Corn ear with tip-concentrated rot. Kernels at the tip are brownish/pinkish-discolored with white-gray mold. Lower ear has healthy yellow kernels.

Reference comparisons:
- Gibberella\_Disease (ref 12): Very strong match — same tip rot pattern, brownish-pink kernels, white mold growth concentrated at tip.
- Gibberella\_Disease (ref 42): Confirms tip-to-base progression pattern.
- Gibberella\_Disease (ref 64): Shows pink discoloration on husks, consistent with the disease.
- Fusarium\_Disease: Scattered patches across the ear, not tip-focused — does not match.
- Diplodia\_Ear\_Rot: Different presentation (dried/bleached), no pink coloration.
- Aspergillus\_Ear\_Rot: Distinctive olive-green mold — clearly different.
- Trichoderma\_Ear\_Rot: Blue-green mold all over — clearly different.

The tip-origin rot with brownish-pink kernel discoloration and white mold is the hallmark of Gibberella ear rot. The visual match to the Gibberella reference (Corn\_Disease\_12.jpg) is strong.
    \end{description}
    \end{tcolorbox}

%% file: traces/trace_appendix_11.tex
\begin{tcolorbox}[
        breakable, colback=gray!3, colframe=gray!50, boxrule=0.5pt,
        title={\footnotesize \textbf{Reasoning Trace:} Corn Diseases — Anthracnose\_Leaf\_Spot\_And\_Top\_Dieback\_\_Corn\_Disease\_47},
        fonttitle=\footnotesize,
    ]
    \footnotesize
    \begin{minipage}[t]{0.65\linewidth}
    \begin{tabular}{@{}ll@{}}
        \textbf{Model:} & opus \\
        \textbf{KB source:} & internet \\
        \textbf{Reference budget (k):} & 8 \\
        \textbf{Prediction:} & Anthracnose\_Leaf\_Spot\_And\_Top\_Dieback \\
        \textbf{Ground truth:} & Anthracnose\_Leaf\_Spot\_And\_Top\_Dieback \\
        \textbf{Outcome:} & \textcolor{green!60!black}{\textbf{Correct}} \\
        \textbf{Confidence:} & 0.95 \\
    \end{tabular}
    \end{minipage}%
    \hfill
    \begin{minipage}[t]{0.30\linewidth}
    \centering
    \fbox{\includegraphics[height=2.5cm]{traces/images/Anthracnose_Leaf_Spot_And_Top_Dieback__Corn_Disease_47.jpg}}\\[2pt]
    {\scriptsize \textit{Test image}}
    \end{minipage}

    \vspace{4pt}
    \hrule
    \vspace{4pt}

    \begin{description}[style=nextline, leftmargin=1.5cm, font=\footnotesize\bfseries]
    \item[Step 1:] \texttt{Observe test image}
    \item[Step 2:] \texttt{View reference: part\_index}
    \item[Step 3:] The test image shows corn stalks with dark black discoloration, particularly at the nodes. The stalks appear broken/split, exposing internal tissue. This is a stem disease.

Stem candidates: Anthracnose, Bacterial\_Stalk\_Rot, Charcoal\_Stalk\_Rot, Common\_Smut, Diplodia\_Stalk\_Rot, Physoderma\_Brown\_Spot, Physoderma\_Stalk\_Rot, Purple\_Leaf\_Sheath, among others.

Top suspects based on symptoms: Physoderma\_Stalk\_Rot (black nodes, snapping), Charcoal\_Stalk\_Rot, Anthracnose (stem), Purple\_Leaf\_Sheath. Let me view references one at a time.
    \item[Step 4:] \texttt{View reference: Corn\_Disease\_17}
    \item[Step 5:] Physoderma Stalk Rot reference shows a green stalk with dark spots, quite different from the test image's dry, heavily blackened stalks. Let me check Charcoal\_Stalk\_Rot next.
    \item[Step 6:] \texttt{View reference: Corn\_Disease\_12}
    \item[Step 7:] Charcoal Stalk Rot shows shredded pith with dark interior. The external dark discoloration is somewhat similar. Let me check Anthracnose stem, Purple\_Leaf\_Sheath, and Diplodia\_Stalk\_Rot.
    \item[Step 8:] \texttt{View reference: Corn\_Disease\_10}
    \item[Step 9:] Anthracnose stem shows shiny black areas on the stalk exterior, similar to the test image. Let me check Purple\_Leaf\_Sheath and Diplodia\_Stalk\_Rot.
    \item[Step 10:] \texttt{View reference: Corn\_Disease\_11}
    \item[Step 11:] Purple\_Leaf\_Sheath shows a green, living plant with brown/dark lesions on the sheath. Very different from the test image's dry, dead stalks. Let me check Diplodia\_Stalk\_Rot.
    \item[Step 12:] \texttt{View reference: Corn\_Disease\_14}
    \item[Step 13:] Diplodia Stalk Rot shows tan/brown dry stalks, not the heavy black discoloration in the test image. Let me check Bacterial\_Stalk\_Rot and Common\_Smut stem.
    \item[Step 14:] \texttt{View reference: Corn\_Disease\_16}
    \item[Step 15:] Bacterial Stalk Rot shows wilted, brown leaves on a still-standing plant. Not similar. Let me check Common\_Smut stem and a second Anthracnose stem reference.
    \item[Step 16:] \texttt{View reference: Corn\_Disease\_176}
    \item[Step 17:] Common Smut shows galls, completely different. Let me view the second Anthracnose stem reference, which was the closest match so far.
    \item[Step 18:] \texttt{View reference: Corn\_Disease\_60}
    \item[Step 19:] This is a very strong match. The Anthracnose stem reference (Corn\_Disease\_60.jpg) shows the exact same presentation as the test image: dry corn stalks laid on wooden boards over rocks, with shiny black discoloration on the rind interspersed with tan/yellow areas, and exposed lighter internal tissue where the stalk is broken. The background, stalk appearance, and black lesion pattern are nearly identical.
    \end{description}
    \end{tcolorbox}

%% file: traces/trace_appendix_12.tex
\begin{tcolorbox}[
        breakable, colback=gray!3, colframe=gray!50, boxrule=0.5pt,
        title={\footnotesize \textbf{Reasoning Trace:} Corn Diseases — Aspergillus\_Ear\_Rot\_\_Corn\_Disease\_19},
        fonttitle=\footnotesize,
    ]
    \footnotesize
    \begin{minipage}[t]{0.65\linewidth}
    \begin{tabular}{@{}ll@{}}
        \textbf{Model:} & opus \\
        \textbf{KB source:} & internet \\
        \textbf{Reference budget (k):} & 8 \\
        \textbf{Prediction:} & Fusarium\_Disease \\
        \textbf{Ground truth:} & Aspergillus\_Ear\_Rot \\
        \textbf{Outcome:} & \textcolor{red}{\textbf{Incorrect}} \\
        \textbf{Confidence:} & 0.55 \\
    \end{tabular}
    \end{minipage}%
    \hfill
    \begin{minipage}[t]{0.30\linewidth}
    \centering
    \fbox{\includegraphics[height=2.5cm]{traces/images/Aspergillus_Ear_Rot__Corn_Disease_19.jpg}}\\[2pt]
    {\scriptsize \textit{Test image}}
    \end{minipage}

    \vspace{4pt}
    \hrule
    \vspace{4pt}

    \begin{description}[style=nextline, leftmargin=1.5cm, font=\footnotesize\bfseries]
    \item[Step 1:] \texttt{Observe test image}
    \item[Step 2:] \texttt{View reference: part\_index}
    \item[Step 3:] The test image shows a corn ear with damage at the tip. Kernels at the tip are discolored, swollen, and some appear damaged with whitish material between them. The rest of the ear looks healthy. This is a seed/ear disease.

Candidates from the seed/pod categories: Aspergillus\_Ear\_Rot, Common\_Smut, Diplodia\_Ear\_Rot, Fusarium\_Disease, Gibberella\_Disease, Trichoderma\_Ear\_Rot, Head\_Smut.

Let me view reference images one per class, starting with top candidates.
    \item[Step 4:] \texttt{View reference: Corn\_Disease\_25}
    \item[Step 5:] Fusarium reference shows whitish mold on scattered kernels mid-ear. The test image has damage concentrated at the tip with swollen kernels. Let me check other candidates.
    \item[Step 6:] \texttt{View reference: Corn\_Disease\_218}
    \item[Step 7:] Common Smut shows large silvery-white galls. Not what's in the test image. Let me check more candidates.
    \item[Step 8:] \texttt{View reference: Corn\_Disease\_15}
    \item[Step 9:] Aspergillus shows olive-green powdery mold. Not matching. Let me check Diplodia, Gibberella, Trichoderma, and Head\_Smut.
    \item[Step 10:] \texttt{View reference: Corn\_Disease\_41}
    \item[Step 11:] Diplodia shows cross-section with grayish-brown rotted kernels. Not matching the test image pattern. Let me check Gibberella.
    \item[Step 12:] \texttt{View reference: Corn\_Disease\_18}
    \item[Step 13:] Gibberella shows pinkish mold at ear tip. The test image doesn't have pink coloration. Let me check Trichoderma and Head\_Smut.
    \item[Step 14:] \texttt{View reference: Corn\_Disease\_26}
    \item[Step 15:] Trichoderma shows dark blue-green conidia. Not matching. Let me check Head\_Smut.
    \item[Step 16:] \texttt{View reference: Corn\_Disease\_139}
    \item[Step 17:] None of these are strong matches. The test image shows tip damage with whitish material, which aligns best with Fusarium description ("whitish to pink fungal growth at ear tip"). Let me use my last view on another Fusarium reference.
    \item[Step 18:] \texttt{View reference: Corn\_Disease\_42}
    \item[Step 19:] This Fusarium reference shows whitish-pink mold on kernels with healthy kernels nearby, similar to the test image. The test image shows:
- Damaged, discolored kernels concentrated at the ear tip
- Whitish material between kernels
- Rest of ear is healthy yellow
- No green, pink, or dark colored mold

This matches Fusarium\_Disease: "whitish to pink fungal growth at ear tip" with "kernels can also be infected without visible symptoms" for the scattered pattern.
    \end{description}
    \end{tcolorbox}

%% file: traces/trace_appendix_13.tex
\begin{tcolorbox}[
        breakable, colback=gray!3, colframe=gray!50, boxrule=0.5pt,
        title={\footnotesize \textbf{Reasoning Trace:} Mango Leaf Disease — Anthracnose\_\_Mango\_Leaf\_105},
        fonttitle=\footnotesize,
    ]
    \footnotesize
    \begin{minipage}[t]{0.65\linewidth}
    \begin{tabular}{@{}ll@{}}
        \textbf{Model:} & sonnet \\
        \textbf{KB source:} & internet \\
        \textbf{Reference budget (k):} & 4 \\
        \textbf{Prediction:} & Anthracnose \\
        \textbf{Ground truth:} & Anthracnose \\
        \textbf{Outcome:} & \textcolor{green!60!black}{\textbf{Correct}} \\
        \textbf{Confidence:} & 0.93 \\
    \end{tabular}
    \end{minipage}%
    \hfill
    \begin{minipage}[t]{0.30\linewidth}
    \centering
    \fbox{\includegraphics[height=2.5cm]{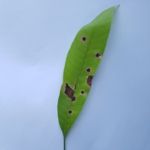}}\\[2pt]
    {\scriptsize \textit{Test image}}
    \end{minipage}

    \vspace{4pt}
    \hrule
    \vspace{4pt}

    \begin{description}[style=nextline, leftmargin=1.5cm, font=\footnotesize\bfseries]
    \item[Step 1:] \texttt{Observe test image}
    \item[Step 2:] The test image shows a mango leaf with discrete dark brown/black spots — irregular, somewhat sunken-looking lesions on the leaf surface. This matches Anthracnose (irregular brown-black spots on leaves) more than the others. Let me check reference images, one per class.
    \item[Step 3:] \texttt{View reference: Mango\_Leaf\_14}
    \item[Step 4:] That anthracnose reference shows a very yellowed/brown leaf — different from the test image. Let me check one from each remaining class.
    \item[Step 5:] \texttt{View reference: Mango\_Leaf\_288}
    \item[Step 6:] \texttt{View reference: Mango\_Leaf\_383}
    \item[Step 7:] This anthracnose reference (Mango\_Leaf\_383) is a near-perfect match to the test image — green leaf with discrete dark brown/black irregular spots scattered across the surface. The sooty mould reference showed a more diffuse dark coating with a greyish cast, not discrete spots.

Summary of evidence:
- Test image: green mango leaf, discrete dark brown-black irregular lesions scattered on surface
- Anthracnose ref 383: identical pattern — green leaf, dark irregular spots, same morphology and distribution
- Sooty mould ref: diffuse dark coating/encrustation, not discrete spots
- Anthracnose KB: "oval or irregular brown to deep brown sunken spots on leaves" — matches exactly
    \end{description}
    \end{tcolorbox}

%% file: traces/trace_appendix_14.tex
\begin{tcolorbox}[
        breakable, colback=gray!3, colframe=gray!50, boxrule=0.5pt,
        title={\footnotesize \textbf{Reasoning Trace:} Mango Leaf Disease — Anthracnose\_\_Mango\_Leaf\_289},
        fonttitle=\footnotesize,
    ]
    \footnotesize
    \begin{minipage}[t]{0.65\linewidth}
    \begin{tabular}{@{}ll@{}}
        \textbf{Model:} & sonnet \\
        \textbf{KB source:} & internet \\
        \textbf{Reference budget (k):} & 4 \\
        \textbf{Prediction:} & Powdery\_Mildew \\
        \textbf{Ground truth:} & Anthracnose \\
        \textbf{Outcome:} & \textcolor{red}{\textbf{Incorrect}} \\
        \textbf{Confidence:} & 0.90 \\
    \end{tabular}
    \end{minipage}%
    \hfill
    \begin{minipage}[t]{0.30\linewidth}
    \centering
    \fbox{\includegraphics[height=2.5cm]{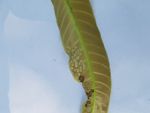}}\\[2pt]
    {\scriptsize \textit{Test image}}
    \end{minipage}

    \vspace{4pt}
    \hrule
    \vspace{4pt}

    \begin{description}[style=nextline, leftmargin=1.5cm, font=\footnotesize\bfseries]
    \item[Step 1:] \texttt{Observe test image}
    \item[Step 2:] The test image shows a mango leaf with white, powdery-looking patches on the surface. This strongly suggests Powdery\_Mildew. Let me check one reference per class.
    \item[Step 3:] \texttt{View reference: Mango\_Leaf\_14}
    \item[Step 4:] \texttt{View reference: Mango\_Leaf\_14}
    \item[Step 5:] \texttt{View reference: Mango\_Leaf\_147}
    \item[Step 6:] \texttt{View reference: Mango\_Leaf\_166}
    \item[Step 7:] Clear differentiation across all four classes:

- Test image: Green leaf with white/cream powdery patches concentrated on the surface
- Powdery\_Mildew ref: White/grey powdery deposits on a green leaf — matches the test image well
- Anthracnose ref: Brown withered leaf with dark necrotic lesions — does not match
- Sooty\_Mould ref: Black soot-like deposits — opposite color (black vs white)
- Die\_Back ref: Completely brown, dried/curled leaf — does not match

The white powdery deposits on an otherwise green leaf are the defining feature of Powdery\_Mildew. No brown lesions (Anthracnose), no black encrustations (Sooty\_Mould), no wilting/browning (Die\_Back).
    \end{description}
    \end{tcolorbox}

%% file: traces/trace_appendix_15.tex
\begin{tcolorbox}[
        breakable, colback=gray!3, colframe=gray!50, boxrule=0.5pt,
        title={\footnotesize \textbf{Reasoning Trace:} Mango Leaf Disease — Anthracnose\_\_Mango\_Leaf\_105},
        fonttitle=\footnotesize,
    ]
    \footnotesize
    \begin{minipage}[t]{0.65\linewidth}
    \begin{tabular}{@{}ll@{}}
        \textbf{Model:} & sonnet \\
        \textbf{KB source:} & no KB \\
        \textbf{Reference budget (k):} & 8 \\
        \textbf{Prediction:} & Anthracnose \\
        \textbf{Ground truth:} & Anthracnose \\
        \textbf{Outcome:} & \textcolor{green!60!black}{\textbf{Correct}} \\
        \textbf{Confidence:} & 0.95 \\
    \end{tabular}
    \end{minipage}%
    \hfill
    \begin{minipage}[t]{0.30\linewidth}
    \centering
    \fbox{\includegraphics[height=2.5cm]{traces/images/Anthracnose__Mango_Leaf_105.jpg}}\\[2pt]
    {\scriptsize \textit{Test image}}
    \end{minipage}

    \vspace{4pt}
    \hrule
    \vspace{4pt}

    \begin{description}[style=nextline, leftmargin=1.5cm, font=\footnotesize\bfseries]
    \item[Step 1:] \texttt{Observe test image}
    \item[Step 2:] Test image: A green mango leaf with several discrete, dark brown/black circular to irregular lesions scattered across the leaf surface. The spots appear sunken or necrotic with defined edges. No powdery coating, no tip dieback, no black sooty film. This pattern strongly suggests Anthracnose.

Let me view one reference from each class before deciding.
    \item[Step 3:] \texttt{View reference: Mango\_Leaf\_14}
    \item[Step 4:] This Anthracnose reference shows a mostly yellowed/brown leaf with extensive necrosis - different from the test image's discrete spots. Let me check other classes.
    \item[Step 5:] \texttt{View reference: Mango\_Leaf\_14}
    \item[Step 6:] \texttt{View reference: Mango\_Leaf\_14}
    \item[Step 7:] \texttt{View reference: Mango\_Leaf\_147}
    \item[Step 8:] Die\_Back is fully dead/shriveled. Powdery\_Mildew shows diffuse white coating. Sooty\_Mould shows dark smudgy patches. The test image has discrete dark spots on a green leaf. Let me check more Anthracnose references.
    \item[Step 9:] \texttt{View reference: Mango\_Leaf\_24}
    \item[Step 10:] This matches the test image closely - green leaf with discrete dark brown/black lesions. Let me view one more Anthracnose and one more Sooty\_Mould to be thorough.
    \item[Step 11:] \texttt{View reference: Mango\_Leaf\_383}
    \item[Step 12:] Excellent match. Let me use my final view on Sooty\_Mould to rule it out.
    \item[Step 13:] \texttt{View reference: Mango\_Leaf\_164}
    \item[Step 14:] Sooty\_Mould shows a diffuse dark film/coating along the leaf surface, not discrete spots. This confirms the test image is not Sooty\_Mould.

Summary:
- Test image: green mango leaf with discrete, well-defined dark brown/black necrotic spots scattered across it
- Anthracnose (Mango\_Leaf\_24, Mango\_Leaf\_383): green leaves with identical discrete dark lesions — strong visual match
- Die\_Back: fully shriveled/brown dead tissue — no match
- Powdery\_Mildew: diffuse white powdery coating — no match
- Sooty\_Mould: dark smudgy film over leaf surface — no match
    \end{description}
    \end{tcolorbox}

%% file: traces/trace_appendix_16.tex
\begin{tcolorbox}[
        breakable, colback=gray!3, colframe=gray!50, boxrule=0.5pt,
        title={\footnotesize \textbf{Reasoning Trace:} Mango Leaf Disease — Anthracnose\_\_Mango\_Leaf\_182},
        fonttitle=\footnotesize,
    ]
    \footnotesize
    \begin{minipage}[t]{0.65\linewidth}
    \begin{tabular}{@{}ll@{}}
        \textbf{Model:} & sonnet \\
        \textbf{KB source:} & no KB \\
        \textbf{Reference budget (k):} & 8 \\
        \textbf{Prediction:} & Sooty\_Mould \\
        \textbf{Ground truth:} & Anthracnose \\
        \textbf{Outcome:} & \textcolor{red}{\textbf{Incorrect}} \\
        \textbf{Confidence:} & 0.72 \\
    \end{tabular}
    \end{minipage}%
    \hfill
    \begin{minipage}[t]{0.30\linewidth}
    \centering
    \fbox{\includegraphics[height=2.5cm]{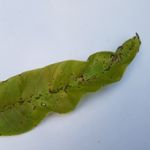}}\\[2pt]
    {\scriptsize \textit{Test image}}
    \end{minipage}

    \vspace{4pt}
    \hrule
    \vspace{4pt}

    \begin{description}[style=nextline, leftmargin=1.5cm, font=\footnotesize\bfseries]
    \item[Step 1:] \texttt{Observe test image}
    \item[Step 2:] The test image shows a mango leaf with small, dark brown/black scattered spots and lesions across the leaf surface. The leaf is green with irregular dark spots, some appearing as small necrotic lesions with irregular margins. No white powdery coating, no sooty black coating, no tip dieback. This pattern suggests Anthracnose.

Let me view one reference per class before deciding.
    \item[Step 3:] \texttt{View reference: Mango\_Leaf\_14}
    \item[Step 4:] This Anthracnose reference shows a yellowed/tan leaf with severe tissue collapse. The test image looks greener with small scattered spots. Let me check Die\_Back.
    \item[Step 5:] \texttt{View reference: Mango\_Leaf\_14}
    \item[Step 6:] Die\_Back shows a completely brown/dead, shriveled leaf - very different from the test image. Let me check Powdery\_Mildew.
    \item[Step 7:] \texttt{View reference: Mango\_Leaf\_14}
    \item[Step 8:] Powdery Mildew shows a grey-white powdery coating across the leaf surface - not matching the test image. Let me check Sooty\_Mould.
    \item[Step 9:] \texttt{View reference: Mango\_Leaf\_147}
    \item[Step 10:] Sooty Mould shows dark scattered patches on a green leaf - somewhat similar to test image. Let me view another Anthracnose reference to compare more carefully.
    \item[Step 11:] \texttt{View reference: Mango\_Leaf\_24}
    \item[Step 12:] Anthracnose\_24 shows larger coalescing dark brown lesions on green leaf. The test image spots are smaller and more scattered. Let me check another Sooty Mould reference.
    \item[Step 13:] \texttt{View reference: Mango\_Leaf\_164}
    \item[Step 14:] This Sooty Mould (164) looks very similar to the test image - wavy green leaf with small scattered dark spots. Let me check one more Sooty Mould and one more Anthracnose.
    \item[Step 15:] \texttt{View reference: Mango\_Leaf\_288}
    \item[Step 16:] \texttt{View reference: Mango\_Leaf\_383}
    \item[Step 17:] Now I have a clear comparison:

- Anthracnose\_383: Larger, well-defined dark brown lesions with clear margins on a green leaf
- Sooty\_Mould\_164: Wavy green leaf with small scattered dark spots - very close to the test image
- Sooty\_Mould\_288: Green leaf with small scattered dark spots and some grey patches

The test image (wavy green leaf, small scattered dark spots) most closely matches Sooty Mould references, particularly Mango\_Leaf\_164. The spots in the test image are smaller and more numerous than typical Anthracnose lesions which tend to be larger with distinct margins.
    \end{description}
    \end{tcolorbox}

%% file: algo.tex
\section{SAGE: Scalable Agentic Grounded Evaluation}

\subsection{Problem Setup}
Let $\mathcal{C} = \{c_1, \dots, c_n\}$ denote the set of crops and $\mathcal{D} = \{d_1, \dots, d_m\}$ denote the set of known diseases. Each disease $d \in \mathcal{D}$ has:

\begin{itemize}
    \item Affected organs: $\mathcal{O}_d \subseteq \mathcal{O}$, where $\mathcal{O}$ is the set of all plant organs (leaf, stem, root, etc.)
    \item Symptom descriptions: $\mathcal{S}_d = \{s_1, s_2, \dots\}$, each with provenance $p(s_i)$ linking to a source quote.
\end{itemize}

Let $\mathcal{I} = \{I_1, \dots, I_q\}$ denote web images. The diagnostic function is defined as:
\[
f_\text{SAGE}: I_\text{test} \times \text{KB} \times \mathcal{R} \rightarrow (d^*, c, \tau)
\]
where $d^*$ is the predicted disease, $c$ is the confidence, and $\tau$ is the reasoning trace.

\subsection{Phase 1: Knowledge Base Curation}

\paragraph{Discovery Agent:} For each $(c,d) \in \mathcal{C} \times \mathcal{D}$, issue web queries:
\[
\mathcal{W}_{c,d} = \text{WebQuery}(c, d)
\]

\paragraph{Extraction Agent:} Extract structured facts from sources:
\[
\mathcal{F}_w = \{(p, t, \mathcal{O}, \mathcal{S})\}, \quad \forall w \in \mathcal{W}_{c,d}
\]
Attach provenance $p(s)$ to each symptom $s \in \mathcal{S}$.

\paragraph{Reconciliation Agent:} Resolve conflicts:
\[
\text{KB} = \text{Reconcile}\Big(\bigcup_{w} \mathcal{F}_w\Big)
\]

\paragraph{Filtering \& Tagging VLM:} For each image $I \in \mathcal{I}$:
\[
\hat{o} = \text{DetectOrgan}(I), \quad
m(I,d) = \text{MatchSymptoms}(I, \mathcal{S}_d)
\]
Keep images where $m(I,d) \ge \theta$ and assign anatomical tags.

\subsection{Phase 2: Agentic Inference}

\paragraph{Observation:}
\[
o_\text{test} = \text{DetectOrgan}(I_\text{test}), \quad
\mathcal{S}_\text{test} = \text{DescribeSymptoms}(I_\text{test})
\]

\paragraph{Narrow Candidates:}
\[
\mathcal{D}_\text{cand} = \{ d \in \mathcal{D} \mid o_\text{test} \in \mathcal{O}_d \}, \quad
\mathcal{D}_\text{rank} = \text{RankBySymptoms}(\mathcal{D}_\text{cand}, \mathcal{S}_\text{test})
\]

\paragraph{Sequential Comparison Loop:} Given reference budget $k$, for $i=1 \dots k$:

\begin{enumerate}
    \item Select candidate disease $d_i = \text{NextCandidate}(\mathcal{D}_\text{rank})$
    \item Fetch reference image $I_\text{ref} = \text{FetchReferenceImage}(\mathcal{R}, d_i, o_\text{test})$
    \item Compare and reason:
    \[
    r_i = \text{CompareAndReason}(I_\text{test}, I_\text{ref}, \mathcal{S}_{d_i})
    \]
    Update reasoning trace $\tau \leftarrow \tau \cup \{r_i\}$
\end{enumerate}

Remove rejected candidates until confident.

\paragraph{Prediction:}
\[
d^* = \arg\max_{d \in \mathcal{D}_\text{rank}} \text{Support}(d, \tau), \quad
c = \text{Confidence}(d^*, \tau)
\]

\begin{algorithm}[H]
\caption{SAGE: Agentic Inference}
\begin{algorithmic}[1]
\Require Test image $I_\text{test}$, KB, Reference set $\mathcal{R}$, Anatomical Index, Reference budget $k$
\Ensure Predicted disease $d^*$, confidence $c$, reasoning trace $\tau$
\State $o_\text{test} \gets \text{DetectOrgan}(I_\text{test})$
\State $\mathcal{S}_\text{test} \gets \text{DescribeSymptoms}(I_\text{test})$
\State $\mathcal{D}_\text{cand} \gets \{d \in \mathcal{D} \mid o_\text{test} \in \mathcal{O}_d\}$
\State $\mathcal{D}_\text{rank} \gets \text{RankBySymptoms}(\mathcal{D}_\text{cand}, \mathcal{S}_\text{test})$
\State $\tau \gets []$
\For{$i = 1$ \textbf{to} $k$}
    \If{Confident($\mathcal{D}_\text{rank}$)}
        \State \textbf{break}
    \EndIf
    \State $d_i \gets \text{NextCandidate}(\mathcal{D}_\text{rank})$
    \State $I_\text{ref} \gets \text{FetchReferenceImage}(\mathcal{R}, d_i, o_\text{test})$
    \State $r_i \gets \text{CompareAndReason}(I_\text{test}, I_\text{ref}, \mathcal{S}_{d_i})$
    \State $\tau.\text{append}(r_i)$
    \If{$r_i$ rejects $d_i$}
        \State $\mathcal{D}_\text{rank}.\text{remove}(d_i)$
    \EndIf
\EndFor
\State $d^* \gets \arg\max_{d \in \mathcal{D}_\text{rank}} \text{Support}(d, \tau)$
\State $c \gets \text{Confidence}(d^*, \tau)$
\State \Return $d^*, c, \tau$
\end{algorithmic}
\end{algorithm}

%% file: references.bib
@article{ahmed2023mangoleafbd,
  title={MangoLeafBD: A comprehensive image dataset to classify diseased and healthy mango leaves},
  author={Ahmed, Sarder Iftekhar and Ibrahim, Muhammad and Nadim, Md and Rahman, Md Mizanur and Shejunti, Maria Mehjabin and Jabid, Taskeed and Ali, Md Sawkat},
  journal={Data in Brief},
  volume={47},
  pages={108941},
  year={2023},
  publisher={Elsevier}
}

@inproceedings{wei2024benchmarking,
  title={Benchmarking in-the-wild multimodal disease recognition and a versatile baseline},
  author={Wei, Tianqi and Chen, Zhi and Huang, Zi and Yu, Xin},
  booktitle={Proceedings of the 32nd ACM International Conference on Multimedia},
  pages={1593--1601},
  year={2024}
}

@article{ghosal2018explainable,
  title={An explainable deep machine vision framework for plant stress phenotyping},
  author={Ghosal, Sambuddha and Blystone, David and Singh, Asheesh K and Ganapathysubramanian, Baskar and Singh, Arti and Sarkar, Soumik},
  journal={Proceedings of the National Academy of Sciences},
  volume={115},
  number={18},
  pages={4613--4618},
  year={2018},
  publisher={National Academy of Sciences}
}

@article{quoc2026leafnet,
  title={LeafNet: A Large-Scale Dataset and Comprehensive Benchmark for Foundational Vision-Language Understanding of Plant Diseases},
  author={Quoc, Khang Nguyen and Dao, Phuong D and Quach, Luyl-Da},
  journal={arXiv preprint arXiv:2602.13662},
  year={2026}
}

@article{hughes2015open,
  title={An open access repository of images on plant health to enable the development of mobile disease diagnostics},
  author={Hughes, David P. and Salath{\'e}, Marcel},
  journal={arXiv preprint arXiv:1511.08060},
  year={2015}
}

@incollection{singh2020plantdoc,
  title={PlantDoc: A dataset for visual plant disease detection},
  author={Singh, Davinder and Jain, Naman and Jain, Pranjali and Kayal, Pratik and Kumawat, Sudhakar and Batra, Nipun},
  booktitle={Proceedings of the 7th ACM IKDD CoDS and 25th COMAD},
  publisher={Association for Computing Machinery},
  pages={249--253},
  year={2020}
}

@misc{ritharson2021sbrd,
  title={Severity-Based Rice Disease Classification},
  author={Ritharson, Isaac},
  year={2021},
  howpublished={\url{https://www.kaggle.com/datasets/isaacritharson/severity-based-rice-leaf-diseases-dataset}},
  note={Kaggle}
}

@misc{marquis2023bean,
  title={Bean Leaf Lesions Classification Dataset},
  author={Marquis03},
  year={2023},
  howpublished={\url{https://www.kaggle.com/datasets/marquis03/bean-leaf-lesions-classification}},
  note={Kaggle}
}

@misc{hayit2020yellowrust,
  title={YellowRust19: Yellow Rust Disease in Wheat},
  author={Hayit, Tolga},
  year={2020},
  howpublished={\url{https://www.kaggle.com/datasets/tolgahayit/yellowrust19-yellow-rust-disease-in-wheat}},
  note={Kaggle}
}

@misc{zarita2022banana,
  title={Banana Leaf Disease Dataset v1.1},
  author={Lozarita, Gimril},
  year={2022},
  howpublished={\url{https://www.kaggle.com/datasets/gimrillozarita/banana-leaf-disease-dataset-v1-1}},
  note={Kaggle}
}

@misc{jstar2023lettuce,
  title={Lettuce Diseases Dataset},
  author={Jena, Ashish},
  year={2024},
  howpublished={\url{https://www.kaggle.com/datasets/ashishjstar/lettuce-diseases}},
  note={Kaggle}
}

@misc{kareem2023cucumber,
  title={Cucumber Plant Diseases Dataset},
  author={Negm, Karim},
  year={2020},
  howpublished={\url{https://www.kaggle.com/datasets/kareem3egm/cucumber-plant-diseases-dataset}},
  note={Kaggle}
}

@misc{cthng2023durian,
  title={Durian Leaf Disease Dataset},
  author={Cthng123},
  year={2025},
  howpublished={\url{https://www.kaggle.com/datasets/cthng123/durian-leaf-disease-dataset}},
  note={Kaggle}
}

@misc{kamalmoha2023eggplant,
  title={Eggplant Disease Recognition Dataset},
  author={Kamalmoha},
  year={2023},
  howpublished={\url{https://www.kaggle.com/datasets/kamalmoha/eggplant-disease-recognition-dataset}},
  note={Kaggle}
}

@misc{basak2023cotton,
  title={Cotton Disease Multi Transformation Dataset},
  author={Basak, Shuvo Kumar},
  year={2026},
  howpublished={\url{https://www.kaggle.com/datasets/shuvokumarbasak2030/cotton-disease-multi-transformation-dataset}},
  note={Kaggle}
}

@misc{basak2023pumpkin,
  title={Pumpkin Leaf Disease Multi Transformation Dataset},
  author={Basak, Shuvo Kumar},
  year={2024},
  howpublished={\url{https://www.kaggle.com/datasets/shuvokumarbasak2030/pumpkin-leaf-disease-multi-transformation-dataset}},
  note={Kaggle}
}

@misc{basak2023rose,
  title={Rose Leaf Disease Multi Transformation Dataset},
  author={Basak, Shuvo Kumar},
  year={2026},
  howpublished={\url{https://www.kaggle.com/datasets/shuvokumarbasak2030/rose-leaf-disease-multi-transformation-dataset}},
  note={Kaggle}
}

@misc{afzaal2023strawberry,
  title={Strawberry Disease Detection Dataset},
  author={Afzaal, Usman},
  year={2021},
  howpublished={\url{https://www.kaggle.com/datasets/usmanafzaal/strawberry-disease-detection-dataset}},
  note={Kaggle}
}

@misc{puspasari2023sugarleafidn,
  title={SugarLeaf-IDN: Sugarcane Leaf Diseases Dataset},
  author={Puspasari, Betty D.},
  year={2026},
  howpublished={\url{https://www.kaggle.com/datasets/bettydpuspasari/sugarleafidn}},
  note={Kaggle}
}

@misc{hayit2022fusarium,
  title={Fusarium Wilt Disease in Chickpea Dataset},
  author={Hayit, Tolga},
  year={2022},
  howpublished={\url{https://www.kaggle.com/datasets/tolgahayit/fusarium-wilt-disease-in-chickpea-dataset}},
  note={Kaggle}
}

@misc{basak2024cauliflower,
  title={Cauliflower Disease Multi Transformation Dataset},
  author={Basak, Shuvo Kumar},
  year={2024},
  howpublished={\url{https://www.kaggle.com/datasets/shuvokumarbasak2030/cauliflower-disease-multi-transformation-dataset}},
  note={Kaggle}
}

@misc{basak2024coconut,
  title={Coconut Disease Multi Transformation STTV Dataset},
  author={Basak, Shuvo Kumar},
  year={2024},
  howpublished={\url{https://www.kaggle.com/datasets/shuvokumarbasak2030/coconut-disease-multi-transformation-sttv-dataset}},
  note={Kaggle}
}

@misc{permana2024vanilla,
  title={Vanilla Plant Disease Image Dataset},
  author={Permana, Muhammad Ihsan},
  year={2024},
  howpublished={\url{https://www.kaggle.com/datasets/mihsanpermana/vanilla-plant-disease-image-dataset}},
  note={Kaggle}
}

@misc{cucumberzenodo2025,
  title={Cucumber Disease and Freshness Classification Dataset --- Curated Annotations},
  year={2025},
  howpublished={\url{https://zenodo.org/records/16816441}},
  note={Zenodo, DOI:~10.5281/zenodo.16816441}
}

@misc{vipoooool2018newplant,
  title={New Plant Diseases Dataset (Augmented)},
  author={{Vipoooool}},
  year={2018},
  howpublished={\url{https://www.kaggle.com/datasets/vipoooool/new-plant-diseases-dataset}},
  note={Kaggle, augmented version of PlantVillage}
}

@misc{rady2024plantimagetext,
  title={Plant Diseases Image-Text Pairs},
  author={{Rady10}},
  year={2024},
  howpublished={\url{https://huggingface.co/datasets/Rady10/Plant-Diseases-Image-Text-Pairs}},
  note={HuggingFace Datasets}
}

@misc{a2h0h0r1plantdiseasenew,
  title={Plant Disease (New) Dataset},
  author={{A2H0H0R1}},
  year={2024},
  howpublished={\url{https://huggingface.co/datasets/A2H0H0R1/plant-disease-new}},
  note={HuggingFace Datasets}
}

@misc{avinashhm2024plantcomplete,
  title={Plant Disease Classification Complete},
  author={{Avinashhm}},
  year={2024},
  howpublished={\url{https://huggingface.co/datasets/avinashhm/plant-disease-classification-complete}},
  note={HuggingFace Datasets}
}

@misc{sakethdevx2024plant,
  title={Plant Disease Dataset},
  author={{Sakethdevx}},
  year={2024},
  howpublished={\url{https://huggingface.co/datasets/sakethdevx/plant-disease-dataset}},
  note={HuggingFace Datasets}
}

@misc{raghavendrad60vqaplant,
  title={VQA Plant-Disease Classification (Merged) Dataset},
  author={{Raghavendrad60}},
  year={2024},
  howpublished={\url{https://huggingface.co/datasets/raghavendrad60/vqa_plant-disease-classification-merged-dataset}},
  note={HuggingFace Datasets}
}

@misc{saon110bdcrop,
  title={Bangladesh Crop \& Vegetable Plant Disease Dataset},
  author={{Saon110}},
  year={2024},
  howpublished={\url{https://huggingface.co/datasets/Saon110/bd-crop-vegetable-plant-disease-dataset}},
  note={HuggingFace Datasets}
}

@misc{bugwood,
  title={Bugwood Image Database System},
  author={{The Bugwood Network and Center for Invasive Species and Ecosystem Health}},
  year={2024},
  howpublished={\url{https://www.bugwood.org/}},
  note={University of Georgia. Per-image attribution required.}
}

@article{pan2025chatleafdisease,
  title={ChatLeafDisease: a chain-of-thought prompting approach for crop disease classification using large language models},
  author={Pan, Jiandong and Zhong, Renhai and Xia, Fulin and Huang, Jingfeng and Zhu, Linchao and Yang, Yi and Lin, Tao},
  journal={Plant Phenomics},
  pages={100094},
  year={2025},
  publisher={Elsevier}
}

@inproceedings{qin2025pdd,
  title={PDD-AGENT: Multimodal Large Language Model-Driven AI Agent for Enhanced Plant Disease Diagnosis},
  author={Qin, Lufu and Wu, Xingcai and Dong, Xinyu and Wang, Huan and Yang, Tingwei and Wang, Qi},
  booktitle={2025 IEEE International Conference on Image Processing (ICIP)},
  pages={1271--1276},
  year={2025},
  organization={IEEE}
}

@inproceedings{liu2024multimodal,
  title={A multimodal benchmark dataset and model for crop disease diagnosis},
  author={Liu, Xiang and Liu, Zhaoxiang and Hu, Huan and Chen, Zezhou and Wang, Kohou and Wang, Kai and Lian, Shiguo},
  booktitle={European Conference on Computer Vision},
  pages={157--170},
  year={2024},
  organization={Springer}
}

@article{zhang2024visual,
  title={Visual large language model for wheat disease diagnosis in the wild},
  author={Zhang, Kunpeng and Ma, Li and Cui, Beibei and Li, Xin and Zhang, Boqiang and Xie, Na},
  journal={Computers and Electronics in Agriculture},
  volume={227},
  pages={109587},
  year={2024},
  publisher={Elsevier}
}

@inproceedings{arshad2025leveraging,
  title={Leveraging vision language models for specialized agricultural tasks},
  author={Arshad, Muhammad Arbab and Jubery, Talukder Zaki and Roy, Tirtho and Nassiri, Rim and Singh, Asheesh K and Singh, Arti and Hegde, Chinmay and Ganapathysubramanian, Baskar and Balu, Aditya and Krishnamurthy, Adarsh and others},
  booktitle={2025 IEEE/CVF Winter Conference on Applications of Computer Vision (WACV)},
  pages={6320--6329},
  year={2025},
  organization={IEEE}
}

@article{zaremehrjerdi2025towards,
  title={Towards large reasoning models for agriculture},
  author={Zaremehrjerdi, Hossein and Ganguly, Shreyan and Rairdin, Ashlyn and Tranel, Elizabeth and Feuer, Benjamin and Di Salvo, Juan Ignacio and Panthulugiri, Srikanth and Pacin, Hernan Torres and Moser, Victoria and Jones, Sarah and others},
  journal={arXiv preprint arXiv:2505.19259},
  year={2025}
}

@inproceedings{awais2025agrogpt,
  title={Agrogpt: Efficient agricultural vision-language model with expert tuning},
  author={Awais, Muhammad and Alharthi, Ali Husain Salem Abdulla and Kumar, Amandeep and Cholakkal, Hisham and Anwer, Rao Muhammad},
  booktitle={2025 IEEE/CVF Winter Conference on Applications of Computer Vision (WACV)},
  pages={5687--5696},
  year={2025},
  organization={IEEE},
}

@article{gauba2025agmmu,
  title={AgMMU: A comprehensive agricultural multimodal understanding and reasoning benchmark},
  author={Gauba, Aruna and Pi, Irene and Man, Yunze and Pang, Ziqi and Adve, Vikram S and Wang, Yu-Xiong},
  journal={arXiv preprint arXiv:2504.10568},
  year={2025}
}

@article{wang2024agri,
  title={Agri-llava: Knowledge-infused large multimodal assistant on agricultural pests and diseases},
  author={Wang, Liqiong and Jin, Teng and Yang, Jinyu and Leonardis, Ales and Wang, Fangyi and Zheng, Feng},
  journal={arXiv preprint arXiv:2412.02158},
  year={2024}
}

@inproceedings{shinoda2025agrobench,
  title     = {{AgroBench}: Vision-Language Model Benchmark in Agriculture},
  author    = {Shinoda, Risa and Inoue, Nakamasa and Kataoka, Hirokatsu and Onishi, Masaki and Ushiku, Yoshitaka},
  booktitle = {arXiv preprint arXiv:2507.20519},
  year      = {2025}
}

@article{zhang2026agri,
  title={Agri-CPJ: A Training-Free Explainable Framework for Agricultural Pest Diagnosis Using Caption-Prompt-Judge and LLM-as-a-Judge},
  author={Zhang, Wentao and Zhang, Qi and Xu, Mingkun and You, Mu and Shen, Henghua and He, Zhongzhi and Jin, Keyan and Wong, Derek F and Fang, Tao},
  journal={arXiv preprint arXiv:2604.23701},
  year={2026}
}

@article{chiranjeevi2025insectnet,
  title     = {{InsectNet}: Real-time identification of insects using an end-to-end machine learning pipeline},
  author    = {Chiranjeevi, Shivani and Saadati, Mojdeh and Deng, Zi K. and Koushik, Jayanth and Jubery, Talukder Z. and Mueller, Daren S. and O'Neal, Matthew and Merchant, Nirav and Singh, Aarti and Singh, Asheesh K. and Sarkar, Soumik and Singh, Arti and Ganapathysubramanian, Baskar},
  journal   = {PNAS Nexus},
  volume    = {4},
  number    = {1},
  pages     = {pgae575},
  year      = {2025},
  month     = jan,
  publisher = {Oxford University Press},
  doi       = {10.1093/pnasnexus/pgae575},
  url       = {https://doi.org/10.1093/pnasnexus/pgae575}
}

@article{shen2025weednet,
  title         = {{WeedNet}: A Foundation Model-Based Global-to-Local {AI} Approach for Real-Time Weed Species Identification and Classification},
  author        = {Shen, Yanben and Ayanlade, Timilehin T. and Boddepalli, Venkata Naresh and Saadati, Mojdeh and Rairdin, Ashlyn and Deng, Zi K. and Arshad, Muhammad Arbab and Balu, Aditya and Mueller, Daren and Singh, Asheesh K. and Everman, Wesley and Merchant, Nirav and Ganapathysubramanian, Baskar and Anderson, Meaghan and Sarkar, Soumik and Singh, Arti},
  journal       = {arXiv preprint arXiv:2505.18930},
  year          = {2025},
  eprint        = {2505.18930},
  archivePrefix = {arXiv},
  primaryClass  = {cs.CV},
  doi           = {10.48550/arXiv.2505.18930},
  url           = {https://arxiv.org/abs/2505.18930}
}
